\documentclass{aastex}
\usepackage{emulateapj5}

\slugcomment{Astrophysical Journal in press}

\shorttitle{$z=7$ LAE}
\shortauthors{Ota et al.}

\begin{document}


\title{Reionization and Galaxy Evolution probed by $z=7$ Ly$\alpha$ Emitters\altaffilmark{1}}

\author{Kazuaki Ota\altaffilmark{2}, Masanori Iye\altaffilmark{3,4,5}, Nobunari Kashikawa\altaffilmark{3,5}, Kazuhiro Shimasaku\altaffilmark{4}, Masakazu Kobayashi\altaffilmark{6}, Tomonori Totani\altaffilmark{6}, Masahiro Nagashima\altaffilmark{7}, Tomoki Morokuma\altaffilmark{3},  Hisanori Furusawa\altaffilmark{8}, Takashi Hattori\altaffilmark{8}, Yuichi Matsuda\altaffilmark{6}, Tetsuya Hashimoto\altaffilmark{4}, Masami Ouchi\altaffilmark{9,10}}
\email{kz\_ota@crab.riken.jp}


\altaffiltext{1}{Based on data collected at Subaru Telescope, which is operated by National Astronomical Observatory of Japan.} 

\altaffiltext{2}{Cosmic Radiation Laboratory, RIKEN, 2-1 Hirosawa, Wako-shi, Saitama 351-0198, Japan}

\altaffiltext{3}{National Astronomical Observatory of Japan, 2-21-1 Osawa, Mitaka, Tokyo, 181-8588, Japan}

\altaffiltext{4}{Department of Astronomy, Graduate School of Science, University of Tokyo, 7-3-1 Hongo, Bunkyo-ku, Tokyo 113-0033, Japan}

\altaffiltext{5}{The Graduate University for Advanced Studies, 2-21-1 Osawa, Mitaka, Tokyo, 181-8588, Japan}

\altaffiltext{6}{Department of Astronomy, Kyoto University, Sakyo-ku, Kyoto 606-8502, Japan}

\altaffiltext{7}{Faculty of Education, Nagasaki University, 1-14 Bunkyo-machi, Nagasaki 852-8521, Japan}

\altaffiltext{8}{Subaru Telescope, 650 North A'ohoku Place, Hilo, Hawaii 96720, USA}

\altaffiltext{9}{Space Telescope Science Institute, 3700 San Martin Drive, Baltimore, Maryland 21218, USA}
\altaffiltext{10}{Hubble Fellow}


\begin{abstract}
We made a narrowband NB973 (bandwidth of 200\AA~centered at 9755\AA) imaging of the 
Subaru Deep Field (SDF) and found two $z=7$ Ly$\alpha$ emitter (LAE) candidates down to NB973 $=24.9$.  
Carrying out deep follow-up spectroscopy, we identified one of them as a real $z=6.96$ LAE.  This has established a new redshift record, showing that galaxy formation was in progress just 750 Myr after the Big Bang.  
Meanwhile, the Ly$\alpha$ line luminosity function of LAEs is known to decline from $z=5.7$ to 6.6 in the SDF.  $L^*$ at $z=6.6$ is 40--60\% of that at $z=5.7$.  We also confirm that the number density of $z=7$ LAE is even only 17\% of the density at $z=6.6$ comparing the latest SDF LAE samples.  This series of significant decreases in LAE density with increasing redshift can be the result of galaxy evolution during these epochs.  
However, using the UV continuum luminosity functions of LAEs, those of Lyman break galaxies 
and a LAE evolution model based on hierarchical clustering, we find that galaxy evolution alone cannot explain all the decrease in density.  This extra density deficit might reflect the attenuation of the Ly$\alpha$ photons from LAEs by the neutral hydrogen possibly left at the last stage of the cosmic reionization at $z \sim 6$--7.
\end{abstract}



%% chosen the first three characterz7LAE_ver2.texs of the first author's name plus

\section{Introduction} 
Investigating high redshift galaxies as well as other distant objects in the early Universe, especially within the first 1 Gyr after the Big Bang, has been the key to understanding how galaxies have formed and evolved, probe their star formation histories, and constrain the epoch of cosmic reionization.  The latest measurements of the polarization of the cosmic microwave background (CMB) by {\it Wilkinson Microwave Anisotropy Probe} ({\it WMAP}) constrained the optical depth to electron scattering during reionization and suggests that the average redshift of reionization was $z=10.9^{+2.7}_{-2.3}$ \citep{Spergel07,Page07}.  Also, Gunn-Peterson (GP) troughs \citep{GP65} in $z \sim 6$ quasar spectra imply reionization ended at $z \sim 6$ with an estimated fraction of intergalactic medium (IGM) neutral hydrogen, $x_{\rm HI}^{z\sim6.2}\sim 0.01$--0.04 \citep{Fan06}.  Moreover, a spectral modeling analysis of a $z\sim6.3$ gamma ray burst (GRB) shows that the Universe seems to have been largely reionized at $z\sim6.3$ with $x_{\rm HI}^{z\sim6.3} = 0$ and the upper limit of $x_{\rm HI}^{z\sim6.3} < 0.17$--0.6, which suggests only some reasonable amount of neutral gas in the GRB host galaxy \citep{Totani06}.  

Another probe of reionization is Ly$\alpha$ emitters (LAEs), young galaxies in the distant universe showing in their spectra redshifted Ly$\alpha$ emission from their interstellar gas illuminated by massive stars.  The observed Ly$\alpha$ line luminosity function (Ly$\alpha$ LF) is expected to decline beyond the redshift $z \sim 6$ where reionization is thought to have completed as the increasing fraction of IGM neutral hydrogen absorbs or scatters the Ly$\alpha$ photons from young galaxies \citep{HiSpa99,RM01,Hu02}.  Nevertheless, recent LAE surveys show that Ly$\alpha$ LF seems not to change from $z = 3$ to 5.7 \citep{Ouch03,Ajiki03,Tran04,vBrk05,Ouch07}.  For the earlier epoch, \citet{MR04} suggest that Ly$\alpha$ LF does not evolve between $z=5.7$ and 6.6.  This might be because their sample could be somewhat biased since it consists of several LAE subsamples taken from various surveys with different kinds of factors such as selection criteria, analysis methods, sky areas, survey volumes, and depths in order to compile as large a sample as possible.  

On the other hand, the Subaru Deep Field \citep[SDF]{kashik04} surveys have tried to keep all these factors as consistent as possible among different redshifts, surveyed exceptionally large volume and made large amount of LAE samples at $z=4.8$, 5.7 and 6.6.  Their latest survey has for the first time confirmed that the Ly$\alpha$ LF declines as $L^*_{z=6.6} \sim L^*_{z=5.7} \times$(0.4--0.6) from $z=5.7$ to 6.6 even after correcting cosmic variance \citep{kashik06}.  From this decline of the LF, they estimated the upper limit of the neutral fraction at $z=6.6$ to be $0 \leq x_{\rm HI}^{z=6.6}\leq 0.45$.  If the nenutral IGM remains at $\sim 50$\% level at $z=6.6$, this constraint supports late reionization and contradicts the {\it WMAP} result.  Also, the decline of Ly$\alpha$ LF at $z=5.7$--6.6 can also be ascribed to the evolution of LAE population itself.

Meanwhile, the ionized fraction $x_i < 1$ and the morphology of H$_{\rm II}$ regions during patchy reionization would modulate the observed distribution of LAEs and enhance the observed clustering of them \citep{Furlanetto06, McQuinn07}.  \citet{McQuinn07} investigated the angular correlation function of the SDF photometric sample of $z=6.6$ LAEs obtained by \citet{kashik06} and suggest that the Universe is fully ionized at $z=6.6$ with the mean volume ionized fraction of $\bar{x_i} \sim 1$.  \citet{McQuinn07} also pointed out the difficulty in distinguishing the effect of evolution of LAE population on Ly$\alpha$ LF from that of reionization.      

LFs of high-$z$ galaxies also tell us about galaxy evolution itself in terms of how many galaxies existed at each luminosity and epoch in the history of the Universe and how it has changed with cosmic time.  To obtain this kind of information at high redshifts, LFs of Lyman break galaxies (LBGs) and LAEs have been mainly observed.  Ultraviolet continuum luminosity functions (UVLFs) of LBGs have been investigated from $z\sim 3$ to $z\sim 7$ and found to decline as redshift increases \citep{LB03,Ouch04,Bouw06,Yoshida06,06BI}.  Since the UV continuum redder than 1216\AA~is not attenuated by neutral IGM hydrogen and if dust extinction is precisely corrected, the decline of UVLF reflects the evolution of galaxies.  

One recently observed example of this is a large decline of UVLF of dropout galaxies at $6 < z \lesssim 7$--8, which is considered to be a clear sign of galaxy evolution over these redshifts \citep{06BI}.  They conclude that very luminous galaxies are quite rare at $z=7$--8.  On the other hand, the UVLF of LAEs was confirmed not to evolve at $z\sim3$--5 \citep{Ouch03}.  In addition, studying LAEs in an even wider sky region, $\sim 1.0$ deg$^2$ of the Subaru/{\it XMM-Newton} Deep Survey (SXDS) field, \citet{Ouch07} found that UVLF of LAEs increases from $z\sim3-4$ to 5.7 while Ly$\alpha$ LF of them remains unchanged over these redshifts, suggesting that the fraction of UV-bright LAEs increases at $z=5.7$.  Furthermore, no evolution of LAE UVLF from $z=5.7$ to 6.6 was observed while Ly$\alpha$ LF of LAEs evolves between these epochs in the latest SDF survey \citep{kashik06}.  This implies that LAEs themselves do not significantly evolve from $z=6.6$ to 5.7 and the decline of the Ly$\alpha$ LF might reflect the effect of reionization.  However, we do not know if this trend of the Ly$\alpha$ LF and UVLF of LAEs continues from even earlier epochs.  In other words, it is not clear whether LAE population evolves from $z>6.6$ as LBG does and the neutral IGM fraction increases to suppress the Ly$\alpha$ LF more severely beyond $z=6.6$.  Moreover, the existence of the galaxies at $z>6.6$ has not been confirmed by spectroscopy yet though several photometric candidates have been found.  These questions can be addressed by observing LAEs and their LFs at $z>6.6$.  Investigating their change over longer cosmic time interval, we can constrain the galaxy evolution and reionization more tightly.

One possible method of detecting $z>6.6$ LAEs is a narrowband filter imaging in infrared region.  However, beyond the redward limit of CCD sensitivity, the large format mosaicing advantages of infrared arrays are not yet available and observations of high redshift LAEs is limited to a small survey volume.  Though recent infrared detectors have achieved extremely high sensitivities, surveys with them cannot avoid large uncertianty due to cosmic variance.  Therefore, we carried out a narrowband survey of $z=7$ LAEs using the final window of OH-airglow at the very edge of the optical regime still accessible with CCDs of Subaru Prime Focus Camera \citep[Suprime-Cam]{Miya02} having a superb wide field of view, $34' \times 27'$.  We chose the wavelength region 9655--9855\AA~open to the highest redshift optical narrowband survey.  Although this might not be considered quite an adequate window since there are several OH lines in the region, the estimated fraction of the sky counts coming from OH lines in the window is not prohibitively large (only $\sim 4.3$ photons s$^{-1}$ \AA $^{-1}$ arcsec$^{-2}$ m$^{-2}$) and we actually succeeded in making a narrowband filter named NB973 covering this wavelength region.  This range corresponds to a redshift of $6.9 \leq z \leq 7.1$ for LAEs.  

To discover such extremely high redshift LAEs and make a sample of them as consistent as possible with those of $z=5.7$ and 6.6 LAEs obtained by \citet{Shima06} and \citet{kashik06}, we targeted the same field, SDF using 8.2-m Subaru Telescope/Suprime-Cam.  Our brief and preliminary result has been recently reported in \citet{06IOK}.  In this survey, we successfully confirmed a $z=6.96$ LAE spectroscopically and observed that the number density of $z=7$ LAE further declines from $z=6.6$ by a factor of 0.18--0.36, suggesting that the neutral hydrogen might increase between these epochs.  However, we do not know whether there had been any possible evolution of LAE population itself from $z=7$ to 6.6, and the density dificit might come from such a galaxy evolution.
       
In this paper, we present the methods and results of our photometric and spectroscopic surveys for $z=7$ LAEs, which were not fully covered in \citet{06IOK}, and try to draw out as much useful information as possible about the epoch of reionization and the LAE galaxy evolution from our results combined with the most recent high redshift galaxy surveys and a LAE evolution model based on hierarchical clustering.  We first describe the properties of our narrowband filter NB973 and imaging observation in \textsection 2.  Then, selection criteria of $z=7$ LAE candidates based on narrow and broadband images are derived in detail and their photometric properties are analyzed in \textsection 3.  In \textsection 4, we explain the results of our follow-up spectroscopy of the selected candidates and their spectroscopic properties.  In \textsection 5, we compare Ly$\alpha$ and UV LFs of $z=7$ LAEs with those of $z=5.7$ and 6.6 LAEs derived from the latest samples obtained by the SDF LAE surveys \citep{Shima06,kashik06} and discuss what implications the result gives for cosmic reionization and galaxy evolution.  Also, any possibilities of LAE galaxy evolution at $z=5.7$--7 are inspected by observational and theoreical approaches.  In the last section, we conclude and summarize our results.  Throughout we adapt a concordance cosmology with ($\Omega_m$, $\Omega_{\Lambda}$, $h$) $= (0.3, 0.7, 0.7)$, and AB magnitudes with $2''$ diameter aperture unless otherwise specified.

\section{Imaging Observation}
\begin{figure}
\plotone{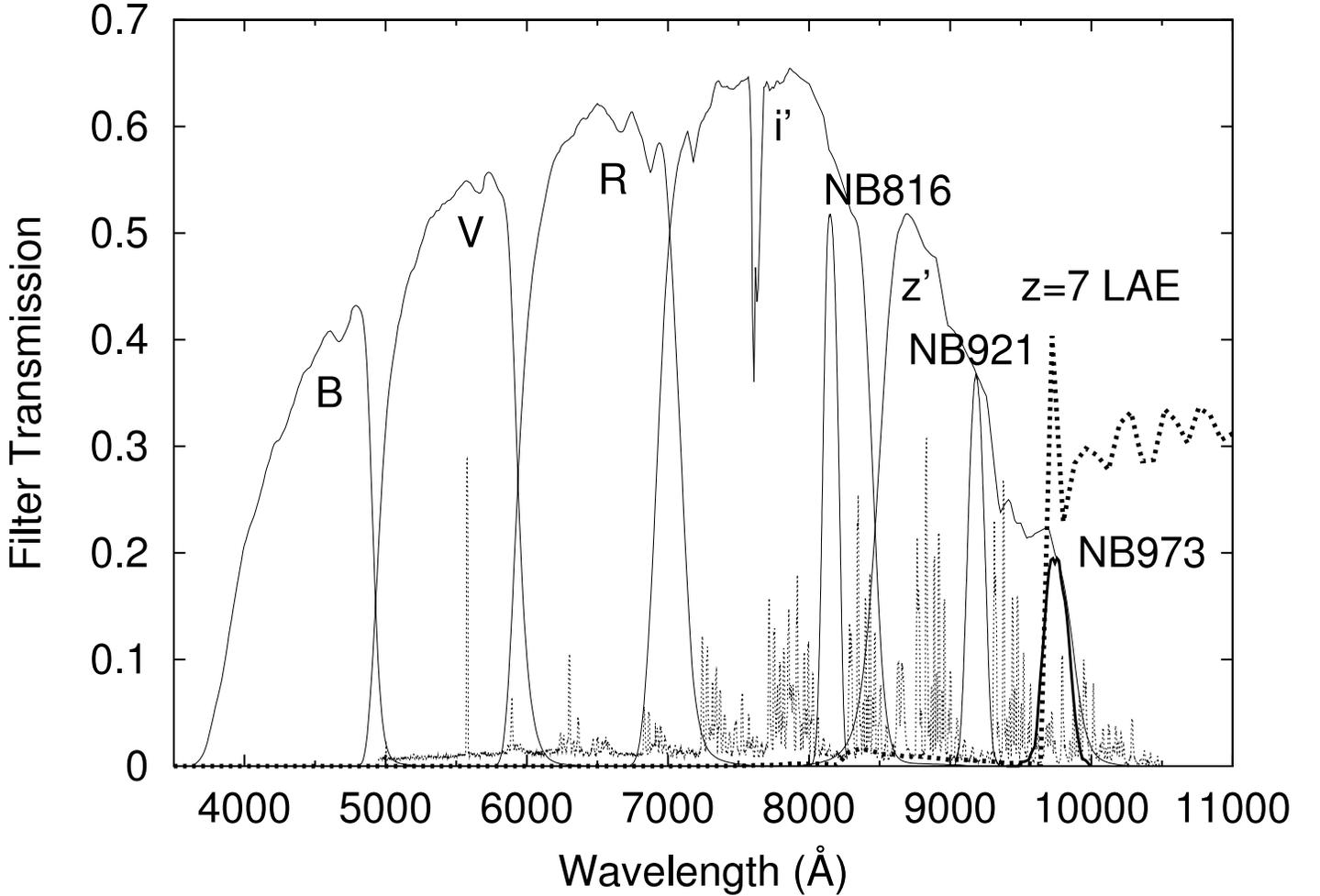}
\caption{Filter transmission of the Suprime-Cam broadbands ($BVRi'z'$: thin solid curves) and narrowbands (NB816, NB921: thin solid curves and NB973: thick solid curve) used for our photometry.  The OH night skylines are also overplotted with thin dashed curve.  Our NB973 filter is $\sim 1.5$ times wider in FWHM than other narrowband filters and includes some OH lines in its bandpass.  The model spectrum energy distribution (SED) of our target, a $z=7$ LAE, obtained using a stellar population synthesis model \citep{BC03} with a metallicity of $Z=Z_{\odot}=0.02$, an age of $t=1$ Gyr, Salpeter initial mass function with lower and upper mass cutoffs of $m_L=0.1$ $M_{\odot}$ and $m_U=100$ $M_{\odot}$ and exponentially decaying star formation history for $\tau=1$ Gyr and with a Ly$\alpha$ emission of the rest frame equivalent width of 50\AA, is also shown with thick dashed line.  The NB973 flux of a $z=7$ LAE is expected to show a strong excess with respect to its $z'$ band flux and should not be detected in other shortward wavebands.\label{MS_BVRizNBfilters_z7LAE}}
\end{figure}

We developed a narrowband filter, NB973, designed to cover the last optical window of OH-airglow centered at 9755\AA~with $\Delta \lambda_{\rm FWHM} \sim 200$\AA~corresponding to Ly$\alpha$ emission at $6.9 \leq z \leq 7.1$ \citep{06IOK}.  The design and fabrication of such a narrowband filter was not a simple issue for Suprime-Cam that uses a fast converging F/1.83 beam whose incident angle varies with a position in the field of view.  Mixture of such light with different incident angles severely degrades the resultant transmission characteristics of the narrowband filter from our target design.  To obtain the desired performance complying with the filter specification for our scientific requirement, we employed a combination of the following three filters glued together: a color cut glass filter RG780 with anti-reflection coating, a narrow bandpass multi-layer coating filter, and another multi-layer coating filter for red leak prevention.  One year before starting the manufacturing of the NB973 filter for Suprime-Cam, we made another filter NB980 (bandwidth of $\sim100$\AA~centered at $\sim9800$\AA) for use in the parallel beam section of the Faint Object Camera And Spectrograph \citep[FOCAS]{kashik02} on Subaru to demonstrate the feasibility of narrowband imaging at this last OH window.  During this prefabrication of NB980, manufacturing errors in controlling the thickness of thin film layers were evaluated.  The multilayer thin film coating design for NB973 was then optimized so that the resulting transmitting properties are relatively robust to inevitable manufacturing errors to control the thickness of each thin layers.  The measured transmission curve of the final NB973 filter actually used in the present survey as well as other filters used for color selection of $z=7$ LAE candidates are shown in Figure \ref{MS_BVRizNBfilters_z7LAE}.

Our target sky region is SDF \citep[13$^{\rm h}$24$^{\rm m}$21.$^{\rm s}$4, -27$^o29'23''$(J2000), $\sim$876 arcmin$^2$]{kashik04}, a blank field in which $z=5.7$ and 6.6 LAE surveys had been also carried out \citep{Shima06, kashik06}.  Deep broadband $BVRi'z'$ and narrowband NB816 ($\lambda_c=8160$\AA, $\Delta\lambda_{\rm FWHM}=$120\AA) and NB921 ($\lambda_c=9196$\AA, $\Delta\lambda_{\rm FWHM}=$132\AA) filter images were taken by the SDF project.  All the images were convolved to have common seeing size of $0.''98$.  Limiting magnitudes in $2''$ aperture at $3\sigma$ are $(B, V, R, i', z', {\rm NB816, NB921})=(28.45, 27.74, 27.80, 27.43, 26.62, 26.63, 26.54)$.  Transmissions of these filters including CCD quantum efficiency, reflection ratio of the telescope prime mirror, the correction for the prime focus optics and transmission to the atmosphere (airmass $\sec z=1.2$) are also shown in Figure \ref{MS_BVRizNBfilters_z7LAE}.

Our NB973 image of the SDF was taken with Suprime-Cam mounted on the Subaru Telescope on 16 and 17 March 2005.  These two nights were photometric with good seeing of $\sim 0.''5$--$0.''8$.  The total integration time is 15 hours.  We have reduced NB973 image frames using the software SDFRED \citep{Ouch04,Yagi02} in the same manner as in \citet{kashik04}.  The NB973 image frames were dithered in a similar way as the SDF project did for other wavebands when they were taken.  The combined NB973 image removed the slight fringing caused by OH-airglow that appeared in some image frames.  The loss of survey area due to this dithering is only $\sim$5\%.  The seeing size of the combined image was $0.''78$ and convolved to $0.''98$, which is the common seeing size of the images of other wavebands, for the purpose of photometry.  Spectrophotometric standard stars Feige34 and Hz44 \citep{Oke90} were imaged during the observation to calibrate the photometric zeropint of the stacked image, which is NB973$=32.03$.  The limiting magnitude reached NB973$\leq24.9$ at $5\sigma$ with 15 hour integration.

%% In this section, we use  the \subsection command to set off
%% a subsection.  \footnote is used to insert a footnote to the text.


\section{Photometric Analysis}

\subsection{Photometry}
After obtaining the stacked NB973 image, we conducted photometric analysis, making an object catalog.  Source detection and photometry were carried out with SExtractor software version 2.2.2 \citep{BA96}.  Pixel size of the Suprime-Cam CCDs is $0.''202$ pixel$^{-1}$.  We considered an area larger than contiguous 5 pixels with a flux [mag arcsec$^{-2}$] greater than $2\sigma$ to be an object.  Object detection was first made in the NB973 image and then photometry was done in the images of other wavebands using the double-imaging mode.  $2''$ diameter aperture magnitudes of detected objects were measured with MAG$_{-}$APER parameter while total magunitudes with MAG$_{-}$AUTO.  Low quality regions of CCDs, bright stellar halos, saturated CCD blooming, and pixels of spiky abnormally high or low flux counts were masked in the SDF images of all wavebands, using the official program code\footnote[11]{Available from http://soaps.naoj.org/sdf/data/} provided by the SDF team \citep{kashik04}.  The final effective area of the SDF image is 876 arcmin$^2$.  The comoving distance along the line-of-sight corresponding to the redshift range $6.94 \leq z \leq 7.11$ for LAEs covered by NB973 filter was 58 Mpc.  Therefore, we have surveyed a total of $3.2 \times 10^5$ Mpc$^3$ volume using NB973 image.  Then, the final object catalog was constructed, detecting 41,533 objects down to NB973$\leq24.9$ ($5\sigma$).    

\subsection{The Detection Completeness}
\begin{figure}
\plotone{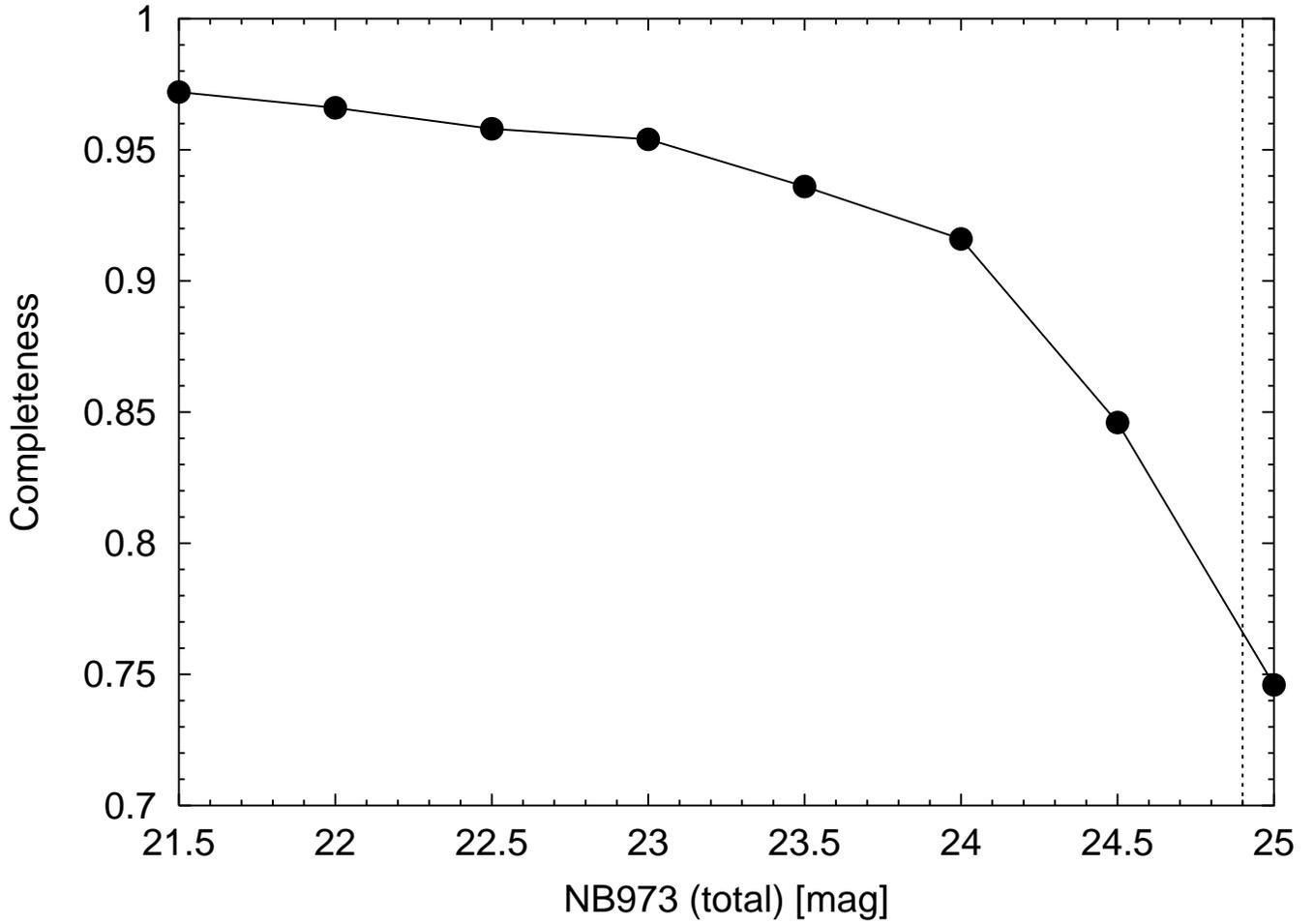}
\caption{The detection completeness of our NB973 image of the SDF, calculated for every 0.5 mag bin.  The dashed line shows our survey limit.  The completeness does not reach 1.0 even for the objects with bright magnitudes.  This is because the blended or overlapped objects tend to be counted as one object by the SExtractor.  The completeness is corrected when the number and luminosity densities of $z=7$ LAE are calculated in \textsection \ref{Re-and-GalEv}.\label{MS_AVE_NB973_Completeness}}
\end{figure}

To understand how reliable our source detections are down to the limiting magnitude of NB973 $\leq 24.9$, we measured the detection completeness of our photometry with the NB973 image.  First, all the objects that satisfy our source detection criterion were removed from the NB973 image using the SExtractor.  Then, the {\tt starlist} task in the {\tt artdata} package of IRAF\footnote[12]{IRAF is distributed by the National Optical Astronomy Observatories, which are operated by the Association of Universities for Research in Astronomy, Inc., under cooperative agreement with the National Science Foundation.} was used to create a sample starlist of about 20,000 artificial objects with a random but uniform spatial and luminosity distributions ranging from NB973 $=20$ to 25 mag.  Next, using the {\tt mkobject} task of IRAF, these artificial objects were spread over the NB973 image, avoiding the masked regions of the SDF and the locations close to the previously removed real objects with the distance shorter than 3/2 of their FWHMs.  After this, SExtractor was run for the source detection in exactly the same way as our actual photometry.  Finally, we calculated the ratios of the number of detected artificial objects to that of created ones to obtain the detection completeness.  We repeated this procedure five times and averaged the obtained completeness.  The result is shown in Figure \ref{MS_AVE_NB973_Completeness}.  The completeness at our detection limit of NB973 $=24.9$ is $\sim76$\%.  The completeness was corrected when the number and luminosity densities of $z=7$ LAEs were calculated in the \textsection \ref{Re-and-GalEv}.

We evaluated the completeness in the same way as \citet{Shima06} and \citet{kashik06} did for the detection completeness of $z=5.7$ and $z=6.6$ LAEs to keep consistency.  However, in real life, some $z=7$ galaxies will lie behind brighter sources at lower redshifts and thus the completeness correction will be artificially small.  The fractions that do can be included in the completeness estimate by simply adding artificial sources to the original image without masking out anything and without any exclusion zones in the placement of artificial sources.  We also calculated our detection completeness in this way to see how much it is different from the original completeness evaluation.  As expected, the completeness we calculated this time is slightly smaller.  However, the difference is a factor of only 1.1--1.3 over NB973 $=20$--25 and it does not change the evaluation of LAE number and luminosity densities much.  Hence, for our consequent analyses, we use the original completesess calculated in the same way as done for $z=5.7$ and $z=6.5$ LAEs to keep consistency.

\subsection{Colors and Selection Criteria of $z=7$ LAEs \label{COLOR_CRITERIA}}
To isolate $z=7$ LAEs from other objects, we investigated their expected colors and derived candidate selection criteria.  We generated model spectra of LAEs at the redshift ranging from $z=5$ to 7 with rest frame Ly$\alpha$ line equivalent width $EW_0(\rm Ly\alpha)$ varying from 0 to 300\AA~as follows.  First, we created a spectral energy distribution (SED) of a starburst galaxy using a stellar population synthesis model, GALAXEV \citep{BC03} with a metallicity of $Z=Z_{\odot}=0.02$, an age of $t=1$ Gyr, Salpeter initial mass function with lower and upper mass cutoffs of $m_L=0.1$ $M_{\odot}$ and $m_U=100$ $M_{\odot}$ and exponentially decaying star formation history for $\tau=1$ Gyr.  These parameters were chosen to be the same as those used to generate model $z=6.6$ LAEs in \citet{Tani05} to keep consistency.  Although recent observational studies show that LAEs seem to be much younger than the 1 Gyr age and/or 1 Gyr star formation decay time and $z=4.5$ LAEs seem to have dust extinction \citep{Gawiser06,Pirzkal07,Finkelstein07}, we did not consider the effects of dust on the SED since the two issues have opposite effects on the broadband colors of the LAEs and this does not have a major effect on the LAE selection criteria.  Then, the SED was redshifted to each of $z=5.0$, 5.5, 5.7, ..., and 7.0, and Ly$\alpha$ absorption by IGM was applied to it, using the prescription of \citet{madau95}.  Finally, flux of a Ly$\alpha$ emission line with either of $EW_0(\rm Ly\alpha)=0$, 10, 20, 50, 100, 150, 200, 250 or 300\AA~was added to the SED at $(1+z)$1216\AA.  We did not assume any specific line profile or velocity dispersion of Ly$\alpha$ emission.  Instead, we simply added 1/2 of the total line flux value, assuming that the blue half of the Ly$\alpha$ line is absorbed by IGM.  An example of a model spectrum of a $z=7$ LAE is shown in Figure \ref{MS_BVRizNBfilters_z7LAE}. 

 %We did not add other emission lines such as H$\beta$, [OIII], [OII], H$\alpha$, [SII] and so on for simplicity because these emissions at lower redshifts can be distinguished from $z=7$ Ly$\alpha$ emission by requiring null flux detection in wavebands shortward of Ly$\alpha$ as stated below.

Colors of these model LAEs were calculated using their SEDs and transmission curves of Suprime-Cam broadband and NB973 filters and plotted in a two-color diagram of $z'-$NB973 vs. $i'-z'$ shown in Figure \ref{MS_2Color_Diagram}.  As clearly seen in the diagram, a $z=7$ LAE is expected to produce significant flux excess in NB973 against $z'$.  However, it should be also noted that NB973 bandpass overlaps with the wavelength range at the longward edge of $z'$ band.  This allows LAEs and LBGs at even lower redshifts $z=6.2$--6.8 to cause the NB973 flux excess with respect to $z'$ band if such galaxies have bright and steep UV continua.  Actually, such objects were detected in our photometry.  Their images and photometric properties are shown and described in Figure \ref{MS_Poststamp_IOK1-5} and \textsection \ref{NB973_EXCESS_OBS}.  Out of these lower redshift galaxies, $z=6.5$--6.6 LAEs can be removed by requiring no detection in the narrowband filter NB921 image whose bandpass corresponds to the Ly$\alpha$ emissions at this redshift range \citep{Koda03,Tani05,kashik06}.  Hence, we classified NB973-excess objects with NB973 $\leq 24.9$ ($5\sigma$, $2''$ aperture), which include our target $z=7$ LAE, into following two categories based on $z'-$ NB973 color.    
\begin{itemize}
\item [(1)] $z=6.9$--7.1 LAEs : $B, V, R, i'$, NB816, $z'$, NB921 $<3\sigma$ 
\item [(2)] $z=6.7$--7.1 LBGs : $B, V, R$, NB816, NB921 $<3\sigma$, $i'-z'>1.3$, $z'-$ NB973 $>1.0$
\end{itemize}              
where $B, V, R, i'$, NB816, $z'$ and NB921 fluxes were measured in total magnitudes while $i'-z'$ and $z'-$NB973 colors in $2''$ aperture magnitudes.  All the $i'$ and $z'$ aperture magnitudes fainter than 27.87 and 27.06 ($2\sigma$ limits), respectively, were replaced by these values in the application of criterion (2).  Since the flux of a LAE shortward of Ly$\alpha$ emission should be absorbed by IGM, no detections ($<3\sigma$) in $B, V, R$, NB816 and NB921 with either red $i'-z'>1.3$ color or no detections in $i'$ and $z'$ were imposed as a part of the criteria.  This can help eliminate interlopers such as L/M/T type dwarf stars and lower redshift galaxies with other type of emission lines (e.g., H$\beta$, [OIII], [OII], H$\alpha$, [SII] and so on).  Also, criterion (1) implies that the robust $z=7$ LAE candidates should show significant excess in NB973 over $z'$ and NB921, $z' -$ NB973 $>1.72$ and NB921 $-$ NB973 $>1.64$.

Note that the color selection criteria (1) and (2) are slightly different from those in \citet{06IOK} in that this time we include null detections in NB816 and NB921 whose bandpasses correspond to Ly$\alpha$ emission at $z=5.65$--5.75 and 6.5--6.6, respectively, to make the criteria more reliable and secure.  In fact, the object IOK-3 detected by \citet{06IOK} satisfied the criteria (1) and (2) simultaneously except for NB921 $< 3\sigma$ and was spectroscopically identified as a $z=6.6$ LAE by \citet{kashik06}.

This time, we found only one object satisfying criterion (1) (hereafter referred to as IOK-1 as in \citet{06IOK}) and none met criterion (2).  In order not to miss faint and diffuse $z=7$ LAEs such as Ly$\alpha$ blobs having extended shapes with fairly bright cores but NB973 $> 24.9$ ($2''$ aperture mag.), we also loosened our detection limit cutoff adopting NB973 $\leq 24.9$ (total mag.) as another limiting magnitude.  This increased the number of objects satisfying criterion (1) by 17 while still no objects fell into (2).  

However, this sample might be contaminated by some spurious objects such as sky residuals and noise due to fringing that might not be removed perfectly at the time of image reduction.  Hence, we visually inspected all the broadband and narrowband images of each color-selected object and only kept those appearing to have condensed and relatively bright cores and excluded those having only diffuse faint shapes with no cores.  More specifically, we removed objects that look apparently artificial such as connected bad pixels, tails of saturated pixels from bright stars and noises of discrete dismembered shapes or pieces of disconnected pixels with fairly large fluxes.  As a result, we were left with one object (hereafter called IOK-2 as in \citet{06IOK}).  The images of IOK-1 and -2 and their photometric properties are shown in Figure \ref{MS_Poststamp_IOK1-5} and Table \ref{Photo-property}, respectively.  The color-magnitude diagram ($z'-$ NB973 vs. NB973) of IOK-1 to -2 as well as all the objects detected down to NB973 $=24.5$ (total mag.) is plotted in Figure \ref{MS_CMD}.  Their two-color diagram is also shown in Figure \ref{MS_2Color_Diagram}.

\subsection{Possibility of Objects with Weak NB973-excess Being $z=7$ LAEs\label{NB973_EXCESS_OBS}}
\begin{figure}
\plotone{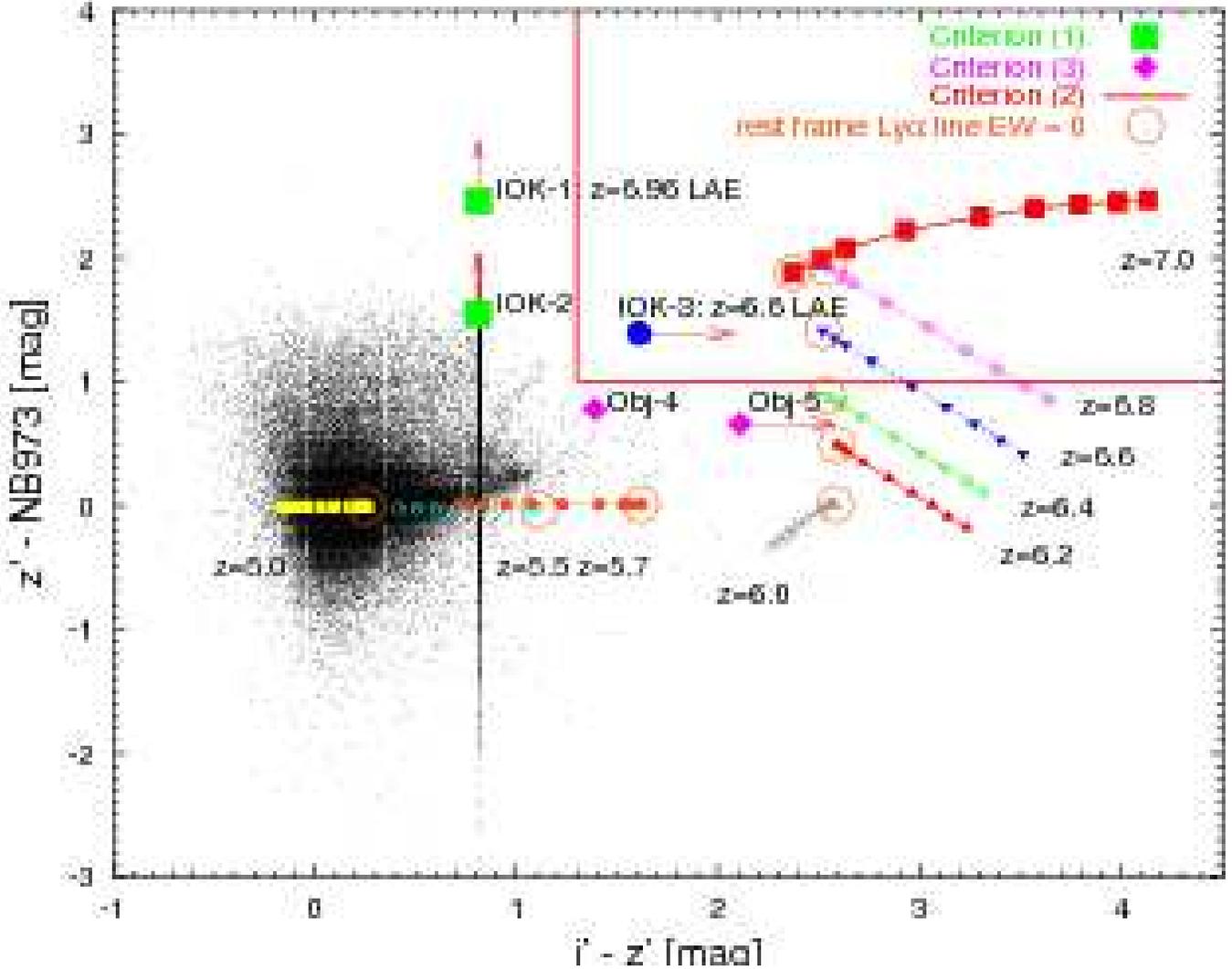}
\caption{$z'-$ NB973 vs. $i'-z'$ plot of the objects with NB973 $\leq 24.9$ (total mag.) detected in the SDF (shown by black dots).  The upper right rectangle region surrounded by the solid line indicates the color selection criterion (2).  Colors and selection criteria of IOK-1, IOK-2 (larger green filled squares) and IOK-3 (larger blue filled circle) as well as Obj-4 and Obj-5 (filled diamonds) are also shown and labeled.  The colors of model LAEs at $z=5.0$--7.0 with the rest frame Ly$\alpha$ line equivalent widths of $EW_0(\rm Ly\alpha)=0, 10, 20, 50, 100, 150, 200, 250$ and 300\AA~are shown in other several different symbols and lines as labeled in the diagram.  Each point with $EW_0(\rm Ly\alpha)=0$, which is the first point in each sequence, is circled.  The model $z=7$ LAEs are smaller red filled squares with a solid line.  All the $i'$ and $z'$ magnitudes fainter than their $2\sigma$ limits were replaced by the $2\sigma$ values in the application of the color selection criteria and are shown by arrows in this diagram.\label{MS_2Color_Diagram}}
\end{figure}

\begin{figure}
\plotone{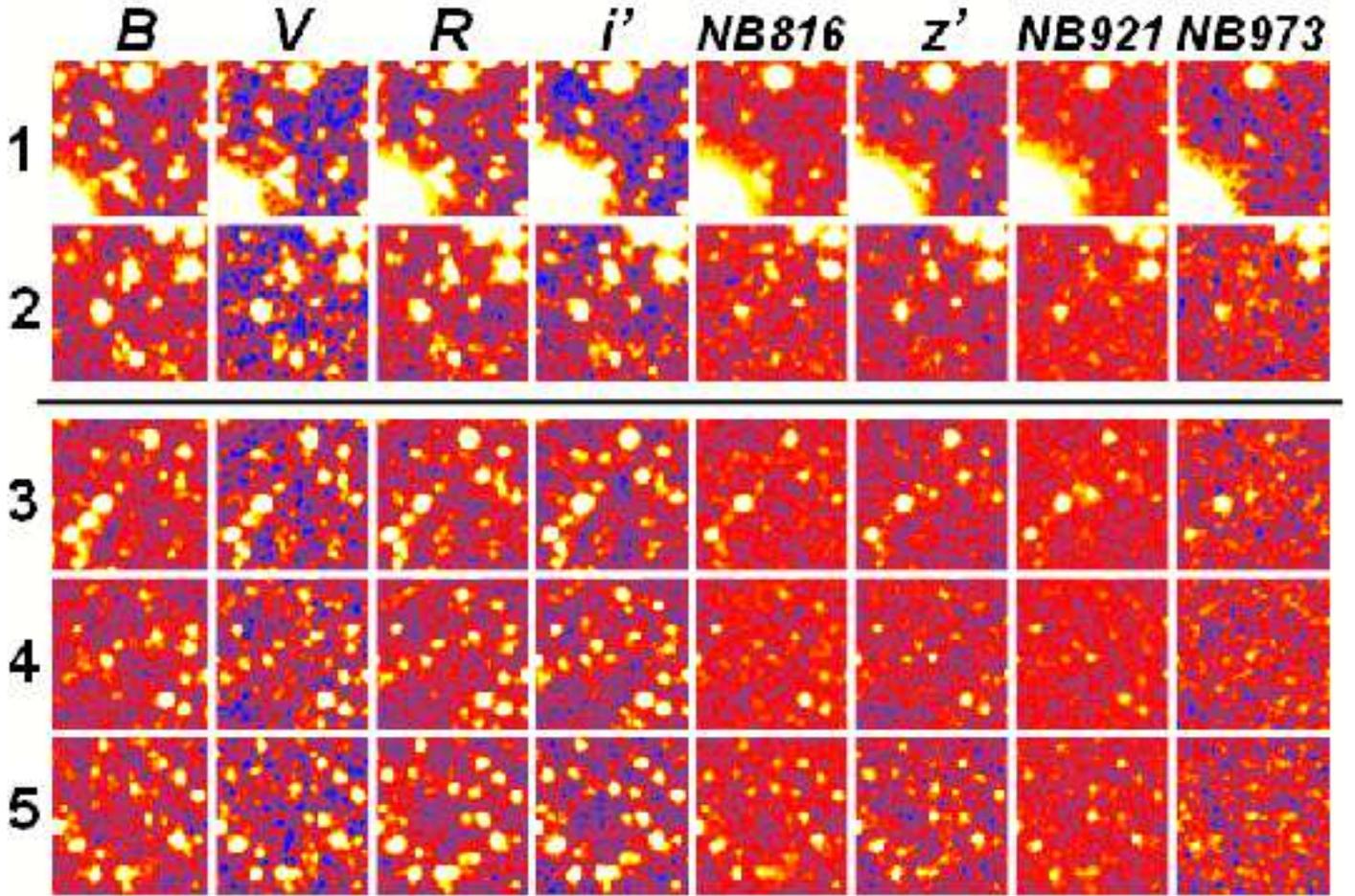}
\caption{The muti-waveband images of the IOK-1, IOK-2 and IOK-3 as well as NB973--3$\sigma$ excess objects, Obj-4 and Obj-5 (labeled by 1, 2, 3, 4 and 5, respectively).  IOK-1, a $z=6.96$ LAE, and IOK-2 are clearly detected only in NB973.  IOK-3, a $z=6.6$ LAE identified by \citet{kashik06}, shows a significant excess in both NB921 and NB973 against $z'$ at the same time but is obviously brighter in NB921.  Obj-4 is seen in all the narrowbands, $i'$ and $z'$ bands while Obj-5 is detected in $z'$, NB921 and NB973.  Both objects show $3\sigma$--excess in $z'-$ NB973 but no excess in NB816 (bandpass for $z=5.65$--5.75 LAEs), NB921 (bandpass for $z=6.5$--6.6 LAEs) and thus could be either of $z=6.2$--6.4 galaxies, low-$z$ EROs or late type dwarf stars.\label{MS_Poststamp_IOK1-5}}
\end{figure}

\begin{figure}
\plotone{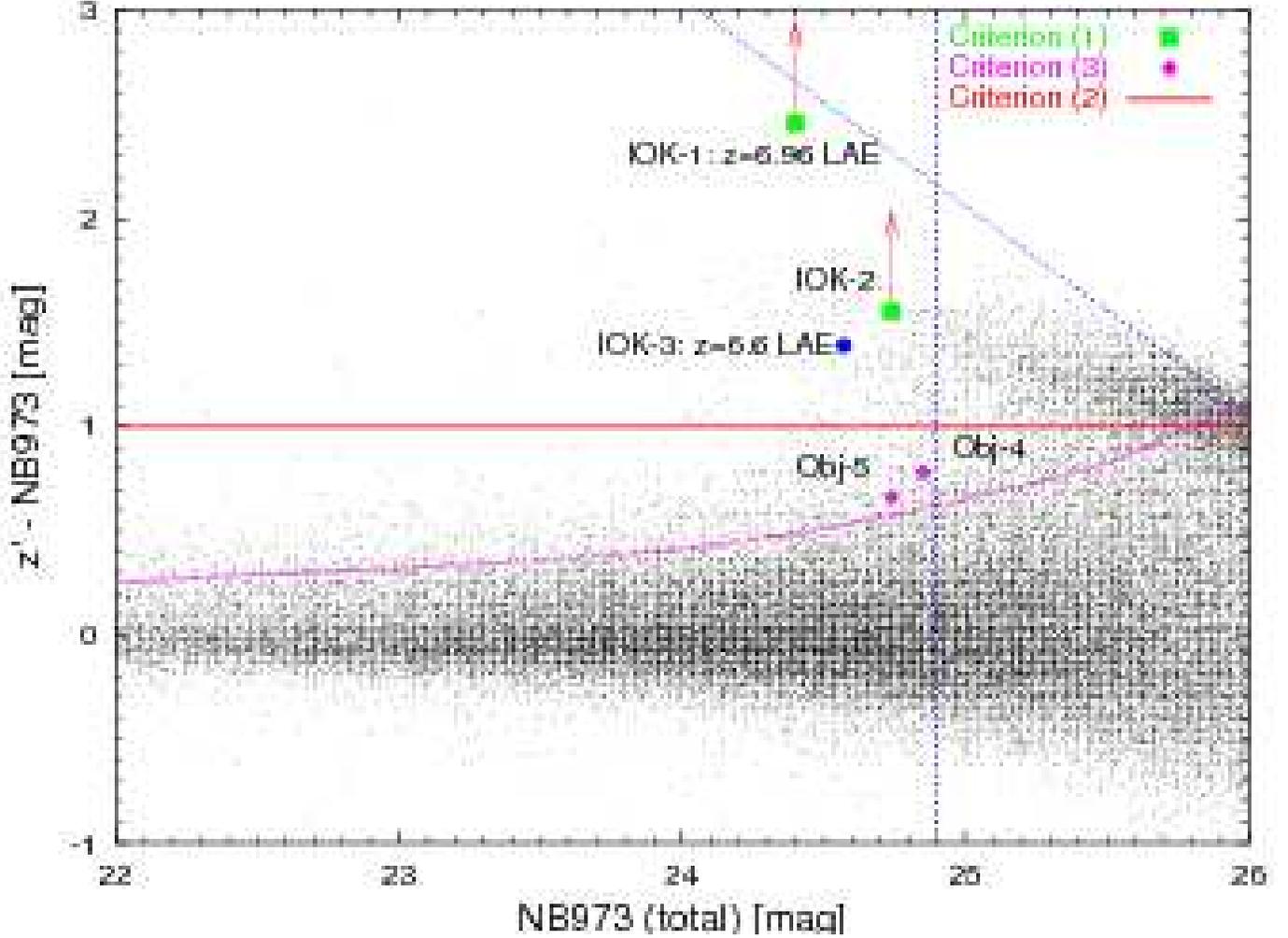}
\caption{$z'-$ NB973 ($2''$ aperture mags.) color as a function of NB973 (total) magnitude of the objects detected in the SDF (shown by dots).  The dotted curve shows $3\sigma$ error track of $z'-$ NB973 color.  The horizontal solid line is a part of our color selection criterion (2), $z'-$ NB973 $>1.0$.  The vertical dashed line indicates the detection limiting magnitude of our survey, NB973 $=24.9 (5\sigma)$.  The diagonal dashed line is the $2\sigma$ limits of $i'$ and $z'$ $2''$ aperture magnitudes.  The IOK-1, -2 and -3 as well as Obj-4 and -5 are denoted by the same symbols as those in Figure \ref{MS_2Color_Diagram}.\label{MS_CMD}}
\end{figure}

  As \citet{Tani05} did in selecting out their candidate $z=6.6$ LAEs in order not to miss faint targets, we also investigated the possibility of objects with a weak excess of $1.0 > z'-$ NB973 $> 3\sigma$ being $z=7$ LAEs even though such objects do not have the expected colors of $z=7$ LAEs predicted by the stellar populationin synthesis model in \textsection \ref{COLOR_CRITERIA} and Figure \ref{MS_2Color_Diagram}.  We define the color criterion of such weak NB973-excess objects as:
\begin{itemize}
\item [(3)] $B, V, R<3\sigma$, $i'-z'>1.3$, $1.0>z'-$ NB973 $>3\sigma$
\end{itemize} 

As mentioned in \textsection \ref{COLOR_CRITERIA}, our NB973 is located at the red edge of $z'$ band.  This could cause the criterion (3) to pick up interlopers such as $z=6.2$--6.8 LAEs/LBGs (as Figure \ref{MS_2Color_Diagram} predicts), $z=1$--3 extremely red objects (EROs) whose continua have the rest frame 4000\AA~Balmer breaks that result in the NB973-excess against $z'$ or M/L/T type red cool dwarf stars whose SEDs can have steep slopes at around NB973 bandpass.  Such objects should be distinguished from $z=7$ LAEs if they are detected in NB973 as well as $z'$.  From the photometry alone, it is difficult to tell if the criterion (3) objects are EROs or dwarfs.  However, it is possible to say whether the objects are galaxies at $z=7$ or not, which is more important in our study. 

According to the predicted colors of model galaxies in Figure \ref{MS_2Color_Diagram}, the criterion (3) should select out $z=6.2$--6.8 LAEs/LBGs, not $z=7$ ones.  However, as some of $z=5.7$ LAEs spectroscopically identified by \citet{Shima06}, though not so many, do not satisfy their color selection criteria computed using SED models, objects satisfying our criterion (3), which reside near the border of criteria (1) and (2), could be $z=7$ LAEs.  We found two objects to fall into criterion (3) (hereafter, referred as to Obj-4, the brighter of the two in $2''$ aperture NB973 mag., and Obj-5).  Their colors, images and photometric properties are shown in Figures \ref{MS_2Color_Diagram}, \ref{MS_Poststamp_IOK1-5}, \ref{MS_CMD} and Table \ref{Photo-property}.  

If they are LAEs, their redshifts can be further constrained by using NB816 and NB921 images.  As seen in Figure \ref{MS_Poststamp_IOK1-5}, Obj-4 is detected in $i'$, NB816, $z'$, and NB921 as well as NB973 but does not show any significant excess in NB816 against $i'$ and in NB921 with respect to $z'$ although it displays NB973-excess greater than $3\sigma$ against $z'$.  Therefore, it is neither a $z=5.65$--5.75 LAE nor $z=6.5$--6.6 one.  Since it is clearly detected in NB816, which is the waveband well shortward of $z=6.7$--7 Ly$\alpha$ emission, Obj-4 could be a $z=6.2$--6.4 LAE or LBG.  

On the other hand, Obj-5 is detected in $z'$ and NB921 as well as NB973 but does not show significant excess in NB921 with respect to $z'$ and thus is not a $z=6.5$--6.6 LAE.  Also, detection in NB921 rules out the possibility of $z=6.7$--7 LAEs since their fluxes shortward of Ly$\alpha$ should be close to zero.  Though it displays an excess of $z'-$ NB973 $>3\sigma$, its $i'-z'$ color is very similar to that of a $z\sim 5.7$ LAE, which is predicted not to produce any NB973-excess.  However, it is not detected in NB816 image and thus not a $z\sim 5.7$ LAE.  Hence, Obj-5 could be a LAE or LBG at $z=6.2$--6.4.  

As mentioned earlier, Obj-4 and -5 can be EROs or dwarfs.  However, we have confirmed that all the objects with weak ($>3\sigma$) NB973-excesses (i.e., Obj-4 and -5) cannot be $z=7$ LAEs and thus do not have to care anymore about the possibility of missing any faint $z=7$ LAE candidates.

\subsection{Possibility of IOK-1 and IOK-2 Being Variable Objects \label{variable}}    
As the selection criterion (1) derived in \textsection \ref{COLOR_CRITERIA} shows, the most probable $z=7$ LAE candidates are imaged in only NB973 waveband and not detected in any of other filters.  Since the NB973 image was taken 1--2 years after the $BVRi'z'$ images of the SDF had been obtained, the sources only bright in NB973 can be some variable objects such as supernovea and active galactic neuclei (AGNs) that accidentally increased their luminosities during our NB973 imaging observation.  Therefore, we investigated how many objects can be such variables.  In another word, this corresponds to the number of the objects that were fainter than our detection limit NB973 $=24.9$ $(5\sigma)$ at some epoch but that can become brigher than it at another epoch.  Since there are no enough data in $z'$ and NB973 bands for the statistic of variables, we instead used $i'$ band images taken over several separate epochs \citep{Kuma07} for the best possible (but somewhat rough) estimation we can do.  

\begin{figure}
\plotone{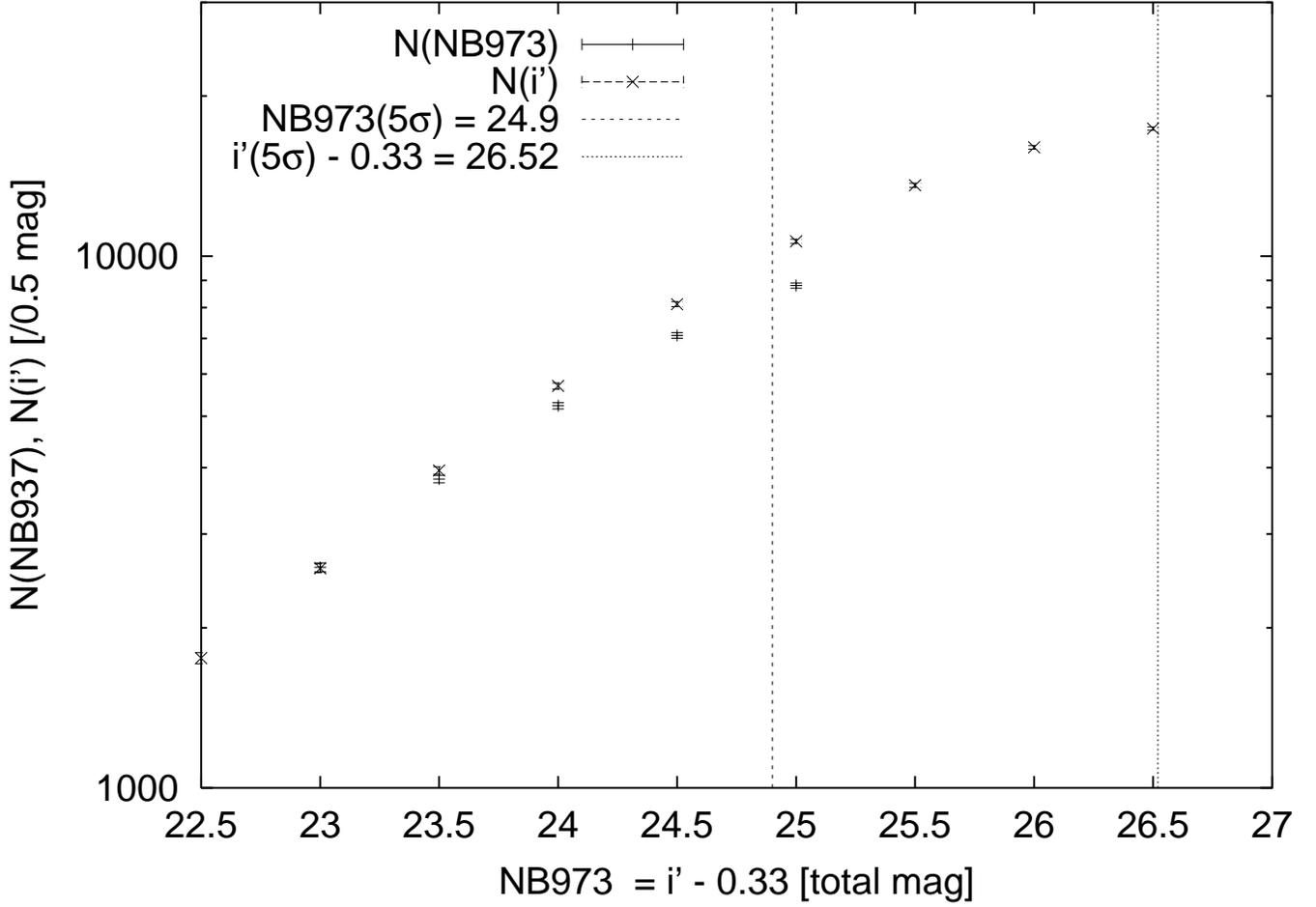}
\caption{Comparison of the number counts of objects detected in the NB973 (plus symbols) and $i'$ band (crosses) images of SDF down to their detection limits, NB973 $=24.9$ $(5\sigma)$ (dashed line) and $i'=26.85$ $(5\sigma)$ (dotted line), respectively.  We try to extrapolate the NB973 number count of objects between NB973 $=24.9$ and 26.52 by using the mean color relation of $<i'-$ NB973$>=0.33$ and $i'$ band number count at the corresponding magnitude range.  Some fraction of such objects can be candidate variables that would become brighter than NB973 $=24.9$ at another occasion.  Since $i'$ band number count is slightly ($\times 1.1$--$\times 1.2$) higher than that of NB973 at NB973 $\leq 24.9$, our extrapolation can be an overestimation.\label{MS_i_Ncount}}
\end{figure}

First, we calculated the mean color of $i'-$ NB973 over the range of NB973 $=$ 22--25, which is $<i'-$ NB973$> = 0.33$, for the purpose of rough conversion of NB973 into $i'$ magnitude.  Using it, NB973 $=24.9$ corresponds to $i'=0.33+24.9=25.23$.  Since the detection limit of SDF $i'$ band image ($i'=$ 26.85 in $5\sigma$) is firmly deeper, the number count of objects fainter than our NB973 detection limit corresponding to $i'=25.23$ can be securely obtained down to $i'=$ 26.85.  The number count per 0.5 mag bin as well as the magnitude increments needed to exceed NB973 $=24.9$ in brightness to be detected in NB973 are shown in Table \ref{N_vs_i}.  Since we were extrapolating the object number counts in NB973 down to NB973 $=26.5$ using $i'$ band object number counts down to $i'= 26.85$ $(5\sigma)$ and $<i'-$ NB973$> = 0.33$, we also checked how similar the number counts in NB973 and in $i'-0.33$ are to each other as shown in Figure \ref{MS_i_Ncount} and Table \ref{N_vs_NB973_or_i}.  Since $i'-0.33$ number counts are slightly larger (by a factor of $\times 1.1$--1.2 per bin), our calculation of the number of variables can be only a little overestimation.  Note that the detection completeness of $i'$ and NB973 are not corrected in their number counts.  This can be the cause of the smaller counts in NB973 than $i'$ toward our detection limit NB973 $=24.9$. 

We use in our calculation four $i'$ images of a part of SDF ($\sim 71$\% of the total area) taken at four separate epochs: 4 March 2005, 30 April 2003, 11 April 2002 and 24 April 2001, respectively \citep{Kuma07}.  The numbers of variable objects $N_v(\Delta i')$ that increased their $i'$ magnitudes by $\Delta i'$ over the periods 2003--2005 and 2001--2002 were counted in each magnitude $\Delta i'$ bin (matched to $\Delta m$ bin in Table \ref{N_vs_i}) as shown in Table \ref{N_Variables}.  In the 2003 and 2005 images, $\sim 70,000$ and 80,000 objects were detected down to their limiting magnitudes $i'=26.3$ and 26.6 ($5\sigma$, $2''$ aperture), respectively.  Similarly, in the 2001 and 2002 images, $\sim 50,000$ and 70,000 objects were detected down to $i'=25.9$ and 26.2 (also, $5\sigma$, $2''$ aperture), respectively.  Thus, taking the averages, we roughly assumed that $N_{obs}=75,000$ and 60,000 objects were detected in 2003--2005 and 2001--2002, respectively and divided the number of variables $N_v(\Delta i')$ by these numbers $N_{obs}$ to obtain the probabilities $P(\Delta i')$ of finding the variables with a brightness increase of $\Delta m = \Delta i'$ in the SDF down to our detection limit.  

Finally, multiplying the probability by the number counts of $i'$-detected objects $N(\Delta m)$ in Table \ref{N_vs_i} and summing all them up, the number of variables that became brighter than NB973 $=24.9$ came out to be $\sim 9$--10.  Note that since the magnitude increse of $0 < \Delta m\leq 0.1$ is really small change and cannot be distinguished from photometric errors in NB973 and $i'$ magnitude measurements, which is also in the order of up to $\sim 0.1$, we ignored the number of variables in $\Delta m=0$--0.1 bin at the time of the summation.  So far, we have considered only the data of the variables that increased their magnitudes over the two epochs and did not treat those having decreased their magnitudes.  If we roughly assume that their numbers are approximately the same, the number of possible variables could be about one half of that we estimated above, which is $\sim 4.5$--5.  Again, this number might be a little overestimation by a factor of $1.1$--1.2 since we for our extrapolation used $i'$ band number count instead of NB973 one, which is smaller as seen in Figure \ref{MS_i_Ncount} and Table \ref{N_vs_NB973_or_i}.  Correcting for this factor, we estimate that the number of variables would be $\sim3.8$--4.5.  This number is slightly different from that reported in \citet{06IOK} since more elaborate calculations were used here.  The estimated number of variables indicates that we cannot completely reject the possibility of narrowband excess of IOK-1 and IOK-2 being due to object variability.  To securely reveal their identities, follow-up spectroscopy of them is required.    

\section{Spectroscopy}
To confirm the reality of our candidate LAEs, IOK-1 and IOK-2, selected by the color selection criteria in \textsection \ref{COLOR_CRITERIA}, we carried out optical spectroscopy of them during 2005--2006 using the Faint Object Camera And Spectrograph \citep[FOCAS]{kashik02} on Subaru.  The observation status is summarized in Table \ref{Spec-Status}.  An Echelle grism (175 lines mm$^{-1}$, resolution $\simeq 1600$) with $z'$ filter and $0.''8$ slit was used to obtain the spectra of 30-min exposure each, dithered along the slit by $\pm1''$.  The spectrum of spectrophotometric standard, either of Feige 34, Feige 110 or BD+28$^{\circ}$4211 \citep{Oke90,Ham94}, was also obtained for each night and used for flux calibration.  The observation data reduction and analysis were all performed in the same manners as in \citet{06IOK}.    

\subsection{IOK-1, a $z=6.96$ Ly$\alpha$ emitter \label{IOK-1Spec}}    
We identified IOK-1, the brighter of the two $z=7$ LAE candidates, as a $z=6.96$ LAE.  The details of the spectroscopic analysis of this object were reported in \citet{06IOK}.  We measured the skewness and weighted skewness of the Ly$\alpha$ emission line in IOK-1 spectrum and obtained $S=0.558\pm 0.023$ and $S_w=9.46\pm 0.39$ \AA, respectively.  See \citet{Shima06} and \citet{kashik06} for the definition of $S$ and $S_w$.  These values show that the line is quite asymmetric and ensure that it is a Ly$\alpha$ emission.  Actually, our $S_w$ value for IOK-1 is comparable to the average weighted skewness of $z=5.7$ and 6.6 LAEs (calculated from the data in \citet{Shima06} and \citet{kashik06}), $<S_w^{z=5.7}>=7.43\pm 1.47$\AA~and $<S_w^{z=6.6}>=7.31\pm 1.51$\AA, respectively.  

The Ly$\alpha$ line flux, $F(\rm Ly\alpha)$, Ly$\alpha$ line luminosity, $L(\rm Ly\alpha)$, the corresponding star formation rate, $SFR(\rm Ly\alpha)$ as well as other spectroscopic properties of the Ly$\alpha$ emission line of IOK-1 are summarized in Table \ref{Spec-Property}.  To estimate $SFR(\rm Ly\alpha)$, we use the following relation derived from Kennicutt's equation \citep{Kenicutt98} with the case B recombination theory \citep{Brockle71}.
\begin{equation}
SFR({\rm Ly\alpha}) = 9.1 \times 10^{-43} L({\rm Ly\alpha}) M_{\odot} {\rm yr}^{-1}
\label{L-to-SFR_conversion}
\end{equation}

In addition, we estimate the UV continuum flux, $F(\rm UV)$, by simply subtracting Ly$\alpha$ emission line flux, $F(\rm Ly\alpha)$, measured in the spectrum from NB973 total flux, $F_{\rm NB973}$, obtained by SExtractor photometry (MAG$_{-}$AUTO).
\begin{equation}
F({\rm UV}) =F_{\rm NB973}-F(\rm Ly\alpha)
\label{F_UV}
\end{equation}
Then, this UV continuum flux can be converted into the UV continuum luminosity, $L_{\nu}(\rm UV)$, and corresponding star formation rate, $SFR(\rm UV)$.  To estimate $SFR(\rm UV)$, we use the following relation \citep{Kenicutt98,madau98}.
\begin{equation}
SFR({\rm UV}) = 1.4 \times 10^{-28} L_{\nu}({\rm UV}) M_{\odot} {\rm yr}^{-1}
\label{L-to-SFR_conversion_UV}
\end{equation}
The spectroscopic properties of the UV continuum of IOK-1 are listed in Table \ref{Spec-Property_UV}.

\subsection{IOK-2\label{IOK-2-Spec}}    
As reported in \citet{06IOK}, although there appears to be an extremely weak emission-like flux at around 9750\AA~($z=7.02$ if this is a Ly$\alpha$ line) within the small gap between OH sky lines, 3 hours integration on the IOK-2 spectroscopy (obtained from 4 May 2005 and 24 April 2006) was not deep enough to confirm if it is real or spurious since we had only S/N $\sim 2$ even though measured within the gap.  This did not allow us to draw any firm conclusion about IOK-2.      
 
To reveal the true entity of this object, we made additional 8 hours follow-up spectroscopy with Subaru/FOCAS on 10 April 2007 (See Table \ref{Spec-Status}).  The seeing during this observing run was $0.''4$--$1''$ with clear sky.  We combined the spectra taken at this night with those obtained in 2006 and 2005 to achieve the total of 11 hours integration.  However, sky-subtracted stacked spectrum has shown neither the emission-like flux at 9750\AA~nor any other spectral features.  We also combined only the spectra taken in 2007 and again could not find any emission lines.  There are no signals that follow the dithering shifts among 30-min spectrum frames.  This result indicates that the extremely weak emission-like flux at 9750\AA~seen in 3 hours stacked spectrum made from 2005 and 2006 frames is spurious.    

To see if our 11 hours spectroscopy has reached the depth required to detect a Ly$\alpha$ emission, we compare the sky background RMS of the stacked spectrum with the Ly$\alpha$ flux calculated from NB973 magnitude of the IOK-2.  If we assume all of the flux in NB973 comes from the Ly$\alpha$ line at $z=7$ and adopt the total magnitude of NB973 $=24.74$ rather than $2''$ aperture one, we obtain the line flux of $F^{\rm phot}({\rm Ly\alpha})=2.9\times 10^{-17}$ erg s$^{-1}$ cm$^{-2}$.  On the other hand, binning of 4 pixels (corresponding to 0.017Mpc at $z=7$) in the spatial direction is used to extract the one dimensional spectrum.   The sky RMS (in terms of flux density) is measured in this spectrum by calculating the variance in unbinned pixels along the dispersion direction within the wavelength range corresponding to NB973 passband 9655--9855\AA, and it is $3.0 \times 10^{-19}$ erg s$^{-1}$ cm$^{-2}$ \AA$^{-1}$.  The FWHM of Ly$\alpha$ line, for example, of $z=6.6$ LAE varies from 5.5 to 14.6\AA~\citep{kashik06,Tani05}.  If we assume the FWHM distribution of $z=7$ LAE is similar, then we obtain the Ly$\alpha$ line flux of $F^{\rm spec}({\rm Ly\alpha})=(1.7$--$4.4)\times 10^{-18}$ erg s$^{-1}$ cm$^{-2}$.  This is 6.6--$17 \times$ fainter than $F^{\rm phot}({\rm Ly\alpha})$, indicating that we have reached enough depth to detect the Ly$\alpha$ line if IOK-2 is a real LAE at $z=7$.  Likewise, even if we use $2''$ aperture magnitude of NB973 $=25.51$, we obtain $F^{\rm phot}({\rm Ly\alpha})=1.4\times 10^{-17}$ erg s$^{-1}$ cm$^{-2}$, and $F^{\rm spec}({\rm Ly\alpha})$ is 3.2--$8.2 \times$ fainter than this.  Furthermore, even if we assume $\sim$ 68\% of the NB973 flux comes from Ly$\alpha$ line as \citet{06IOK} did, $F^{\rm spec}({\rm Ly\alpha})$ is still 4.5--12 (2.2--$5.6)\times$ fainter than $F^{\rm phot}({\rm Ly\alpha})$ if we use NB973 total ($2''$ aperture) magnitude to calculate $F^{\rm phot}({\rm Ly\alpha})$.  In all cases we have considered, our spectroscopy reached enough depth to detect a Ly$\alpha$ line.  Hence, IOK-2 might not be a LAE.  However, we should note that residuals of the subtracted OH skylines around 9790\AA~in the 11 hours stacked spectrum is still locally strong ($\sim 13$\% of the NB973 pass band is contaminated) and a Ly$\alpha$ line can be masked out if it is weak and redshifted there.     

If IOK-2 is not a LAE, the possible origin of the NB973 flux excess can be one of either a LBG at $z\sim7$, a low-$z$ ERO, a late-type star, a variable object or a noise.  In the former three cases, spectroscopy could show no signals in the spectrum if their continuum light is very faint.  If IOK-2 is a variable object, well possible as discussed in section \ref{variable}, it could have been fainter than our detection limit at the time of the follow-up spectroscopy.  The possibility of IOK-2 being a noise spike in NB973 image cannot be ruled out though it is very low as described in \citet{06IOK}.  An additional NB973 imaging of SDF will be helpful to see if IOK-2 is either of a variable object or a noise.  For the statistical study in the following sections, we hereafter consider that only IOK-1 is a $z=7$ LAE we have successfully identified and IOK-2 is not.

\section{Implications for the Reionization and Galaxy Evolution \label{Re-and-GalEv}}
From $z\sim 6$ quasar GP daigonostics, the neutral IGM fraction at this redshift was estimated to be $x_{\rm HI}^{z\sim 6.2}\sim 0.01$--0.04 and thus the reionization is believed to have already completed at around this epoch \citep{Fan06}.  This result is also supported by the spectral modeling analysis of the currently most distant GRB at $z\sim 6.3$ conducted by \citet{Totani06}, placing the constraint of $0 \leq x_{\rm HI}^{z \sim 6.3} < 0.17$--0.6.  

On the other hand, the observed Ly$\alpha$ LFs of LAEs at $z\sim6$ and higher redshifts can be used to probe the epoch of reionization.  The Ly$\alpha$ LF is expected to decline beyond $z\sim6$ due to a rapid change of neutral IGM and ionization states before and after the completion of the reionization \citep{HiSpa99,RM01,Hu02}.  While the Ly$\alpha$ LFs have been observed not to evolve at $z=3$--5.7 \citep{Ouch03,Ajiki03,Tran04,vBrk05}, it was recently found to decline as $L^*_{z=6.6} \sim L^*_{z=5.7} \times (0.4$--0.6) from $z=5.7$ to 6.6 in SDF suggesting that $x_{\rm HI}^{z=6.6}\leq 0.45$ \citep{kashik06}.  Furthermore, we also found that the number density of $z=7$ LAEs is only 18--36\% of the density at $z=6.6$ \citep{06IOK}.  This series of decrements in densities might reflect the completion of reionization at around $z\sim6$, beyond which the fraction of the neutral IGM hydrogen could possibly increase and attenuate the Ly$\alpha$ photons from LAEs.  

However, this interpretation was based on the assumption that there had been no evoluion of LAE population from $z=5.7$ to 7.  The recent photometric study of $z\sim 6$ $i$-dropouts and $z\sim 7$--8 $z$-dropouts in the Hubble Ultra Deep Field (UDF) demonstrated that galaxy number density decreases by a factor of $\sim 0.1$--0.2, suggesting the rapid evolution of luminous galaxies between these epochs \citep{06BI}.  

In the following discussion, we re-evaluate the comparison of our LAE number and Ly$\alpha$ luminosity densities at $z=7$ with those at $z=5.7$ and 6.6, using the most up-to-date SDF data from \citet{Shima06} and \citet{kashik06}.  We also investigate the possibility of LAE galaxy evolution between $z=5.7$ and 7 and the degree to which it contributed to the number density deficit between these epochs. 

%\subsection{Implication for the Reionization \label{Reionization}}
\subsection{The Evolution of Ly$\alpha$ LF at $z \gtrsim 6$ \label{Reionization}}
\begin{figure}
\plotone{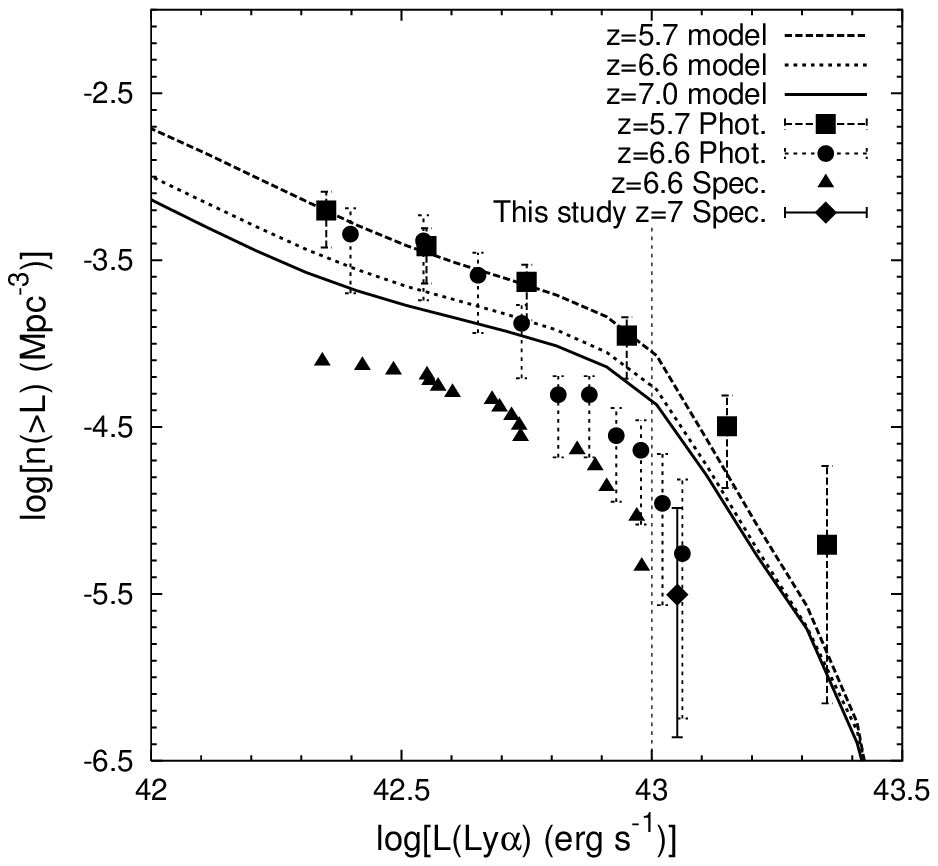}
\caption{Cumulative Ly$\alpha$ LFs of LAEs at $z=5.7$ \citep[filled squares for their photmetric LF]{Shima06}, 6.6 \citep[filled circles and triangles for their photometric (upper limit) and spectroscopic (lower limit) LFs, respectively]{kashik06} and 7 (this study, the diamond for IOK-1 spectroscopic data).  All the errors include cosmic variance and Poissonian errors for small-number statistics (See \textsection \ref{Reionization} for details).  All the data and errors are corrected for their detection completeness.  The long-dashed, short-dashed and solid curves are the intrinsic ({\it i.e.}, not affected by neutral IGM) Ly$\alpha$ LFs at $z=5.7$, 6.6 and 7, respectively, predicted by K07 LAE evolution model (See \textsection \ref{Model_GalEvol}).  The vertical dashed line shows our survey limit to which the LFs are integrated down to obtain LAE number and luminosity densities.\label{MS_Kobayashi-LyaLFs}}
\end{figure}

\begin{figure}
\plotone{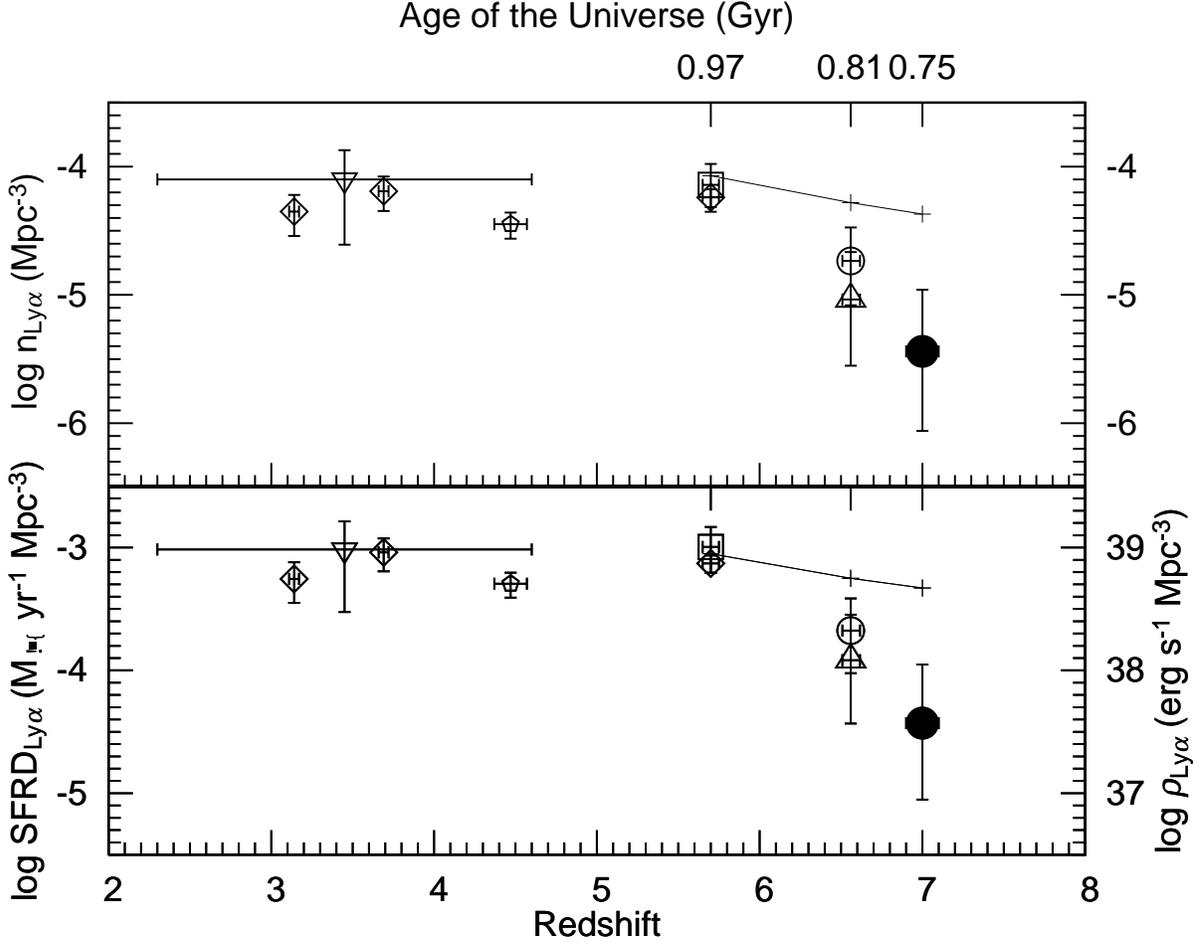}
\caption{The number density $n_{\rm Ly\alpha}$, Ly$\alpha$ line luminosity density $\rho_{\rm Ly\alpha}$ and star formation rate density $SFRD_{\rm Ly\alpha}$ of LAEs at $z=5.7$, 6.6 and 7 derived from the latest Subaru/Suprime-Cam LAE surveys and those at $2.3 < z < 5.7$ from literature down to $L_{\rm limit}(\rm Ly\alpha)=1.0\times 10^{43}$ erg s$^{-1}$.  The densities at $z=5.7$, 6.6 and 7 in the SDF are calculated from the photometric sample of \citet[open squares]{Shima06}, the photometric and spectroscopic samples of \citet[open circles and triangles, respectively]{kashik06}, and IOK-1 spectrum (large filled circles), respectively.  Densities at $2.3 < z < 4.5$ and $z\sim 4.5$ are calculated using the best-fit Ly$\alpha$ Schechter LFs from \citet[inverse triangle]{vBrk05} and \citet[pentagon]{Dawson07}, respectively.  Also, densities at $z=3.1$, 3.7 and 5.7 in the $\sim 1.0$ deg$^2$ of SXDS field are calculated using the best-fit Ly$\alpha$ Schechter LFs from \citet[diamonds]{Ouch07}  Each horizontal error bar shows the redshift range of each survey.  The vertical error bars at $z=5.7$, 6.6 and 7 include both cosmic variance and Poissonian errors for small-number statistics  while those at $z<5.7$ contain only cosmic variance since Poissonian errors for them are negligibly small (See \textsection \ref{Reionization} for details).  The data and vertical error bars at $z=5.7$, 6.6 and 7 are corrected for their detection completeness.  The plus symbols at $z=5.7$, 6.6 and 7 with solid lines show the expected densities obtained by integrating the intrinsic ({\it i.e}, not affected by neutral IGM) Ly$\alpha$ LFs predicted by the K07 LAE evolution model in \textsection \ref{Model_GalEvol}.  At $z>5.7$, the densities clearly decrease with increasing redshifts and smaller than the model-predicted values, implying that the Ly$\alpha$ lines might be attenuated by the possibly increasing neutral IGM at the reionization epoch.\label{MS_Madau_PLOT}}
\end{figure}

Figure \ref{MS_Kobayashi-LyaLFs} compares Ly$\alpha$ line LFs at $z=5.7$, 6.6 and 7 derived from the latest SDF data (that is, \citet{Shima06}, \citet{kashik06}, and \citet{06IOK}).  In addition, Figure \ref{MS_Madau_PLOT} shows the LAE number densities, $n_{\rm Ly\alpha}$, Ly$\alpha$ line luminosity densities, $\rho_{\rm Ly\alpha}$, and corresponding star formation rate densities, $SFRD_{\rm Ly\alpha}$, at $2.3 < z \leq 7$ down to our detection limit $L_{\rm limit}(\rm Ly\alpha) = 1.0 \times 10^{43}$ erg s$^{-1}$ (converted from NB973 $\leq 24.9$ ($5\sigma$) as \citet{06IOK} did).  $\rho_{\rm Ly\alpha}$ and $SFRD_{\rm Ly\alpha}$ at $z=7$ are calculated using Ly$\alpha$ line luminosity estimated from the spectrum of IOK-1 and equation \ref{L-to-SFR_conversion}.  The number and luminosity densities at $z < 7$ are obtained by integrating the best-fit Ly$\alpha$ Schechter LFs \citep{Schechter76} of LAEs down to our detection limit $L_{\rm limit}(\rm Ly\alpha)$ as follows.      
\begin{equation}
\phi(L)dL=\phi^*\left(\frac{L}{L^*}\right)^{\alpha}\exp\left(\frac{-L}{L^*}\right)d\left(\frac{L}{L^*}\right)
\label{Scheter-LF}
\end{equation}
\begin{equation}
n_{\rm Ly\alpha}=\int_{L_{\rm limit}}^{\infty}\phi(L)dL
\label{integ-Scheter-LF}
\end{equation}
\begin{equation}
\rho_{\rm Ly\alpha}=\int_{L_{\rm limit}}^{\infty}\phi(L)LdL
\label{integ-L*Scheter-LF}
\end{equation}

We adopt $(\log(\phi^* [$Mpc$^{-3}]), \log(L^* [$erg s$^{-1}]), \alpha)=(-3.44^{+0.20}_{-0.16}, 43.04^{+0.12}_{-0.14}, -1.5)$ and $(-2.88^{+0.24}_{-0.26}$, $42.60^{+0.12}_{-0.10}, -1.5)$ for  Ly$\alpha$ LFs of LAEs at $z=5.7$ and 6.6 in the SDF (taken from Table 3 in \citet{kashik06}), respectively.  $\phi ^* = (22.0 \pm 12.0, 1.7 \pm 0.2) \times 10^{-4} {\rm Mpc}^{-3}$ and $L^*= (5.4 \pm 1.7, 10.9 \pm 3.3) \times 10^{42}$ erg s$^{-1}$ with $\alpha=-1.6$ are quoted for $2.3<z<4.5$ \citep{vBrk05} and $z\sim4.5$ \citep{Dawson07} Ly$\alpha$ LFs, respectively.  We also use $\phi ^* = (9.2^{+2.5}_{-2.1}, 3.4^{+1.0}_{-0.9}, 7.7^{+7.4}_{-3.9}) \times 10^{-4} {\rm Mpc}^{-3}$ and $L^*= (5.8^{+0.9}_{-0.7}, 10.2^{+1.8}_{-1.5}, 6.8^{+3.0}_{-2.1}) \times 10^{42}$ erg s$^{-1}$ with $\alpha=-1.5$ for Ly$\alpha$ LFs of LAEs at $z=3.1$, 3.7 and 5.7 in $\sim 1.0$ deg$^2$ of the SXDS field \citep{Ouch07}.  $\rho_{\rm Ly\alpha}$ is converted to $SFRD_{\rm Ly\alpha}$ by using equation \ref{L-to-SFR_conversion}.  

\begin{figure}
\plotone{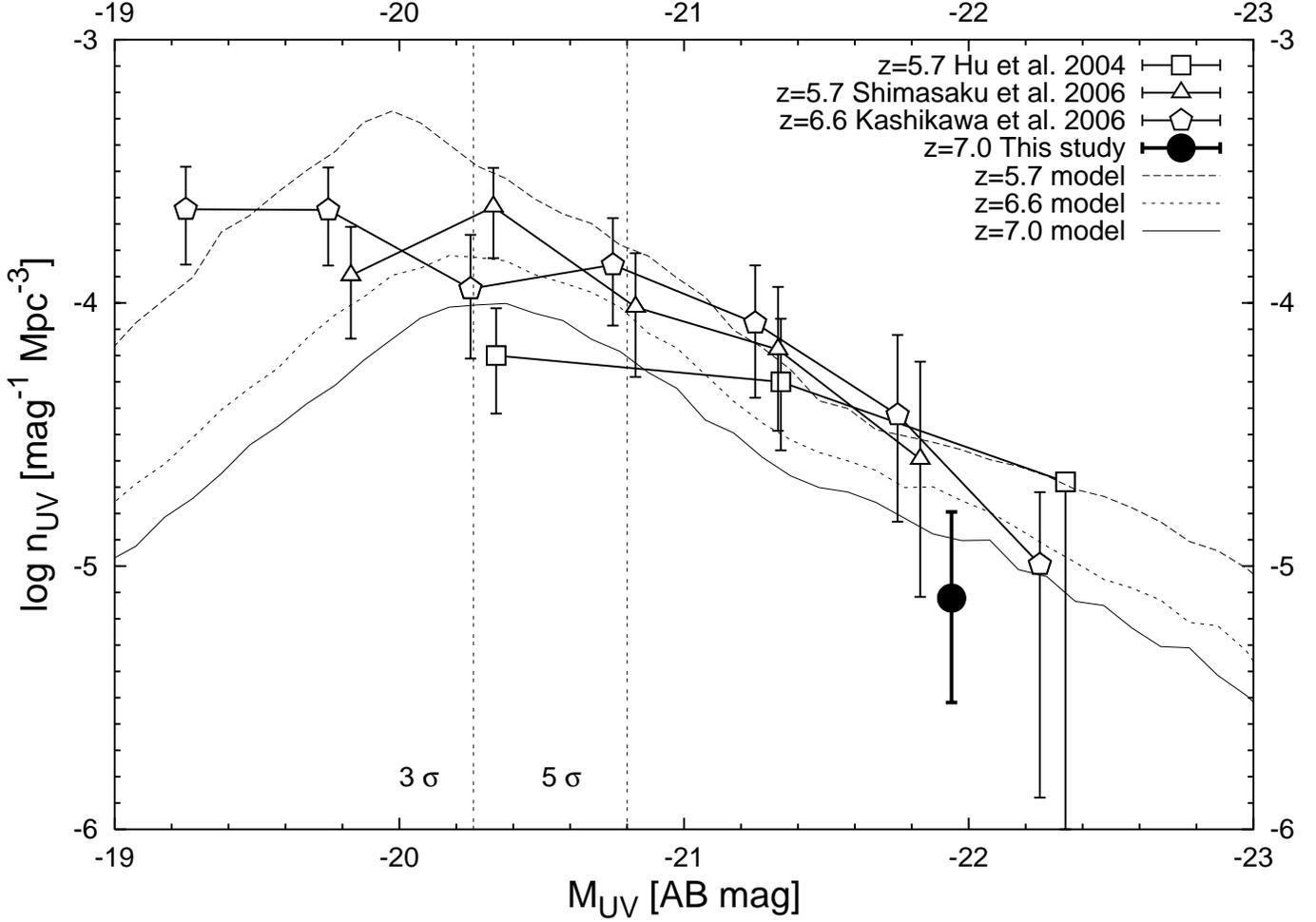}
\caption{The rest frame UVLFs (per unit absolute magnitude) of $z=5.7$ and 6.6 LAEs compared with that at $z=7$ derived using IOK-1 data.  Our survey detection limits at $5\sigma$ and $3\sigma$ are shown by vertical dashed lines.  The errors include cosmic variance and Poissonian errors for small-number statistics (See \textsection \ref{Reionization} for details).  All the data and errors are corrected for their detection completeness.  The $z=7$ UVLF does not evolve from $z=5.7$--6.6.  The long-dashed, short-dashed and solid curves are UVLFs of LAEs at $z=5.7$, 6.6 and 7 predicted by K07 model, respectively.  They are in good agreement with the observation data.\label{MS_UVLF}}  
\end{figure}

\begin{figure}
\plotone{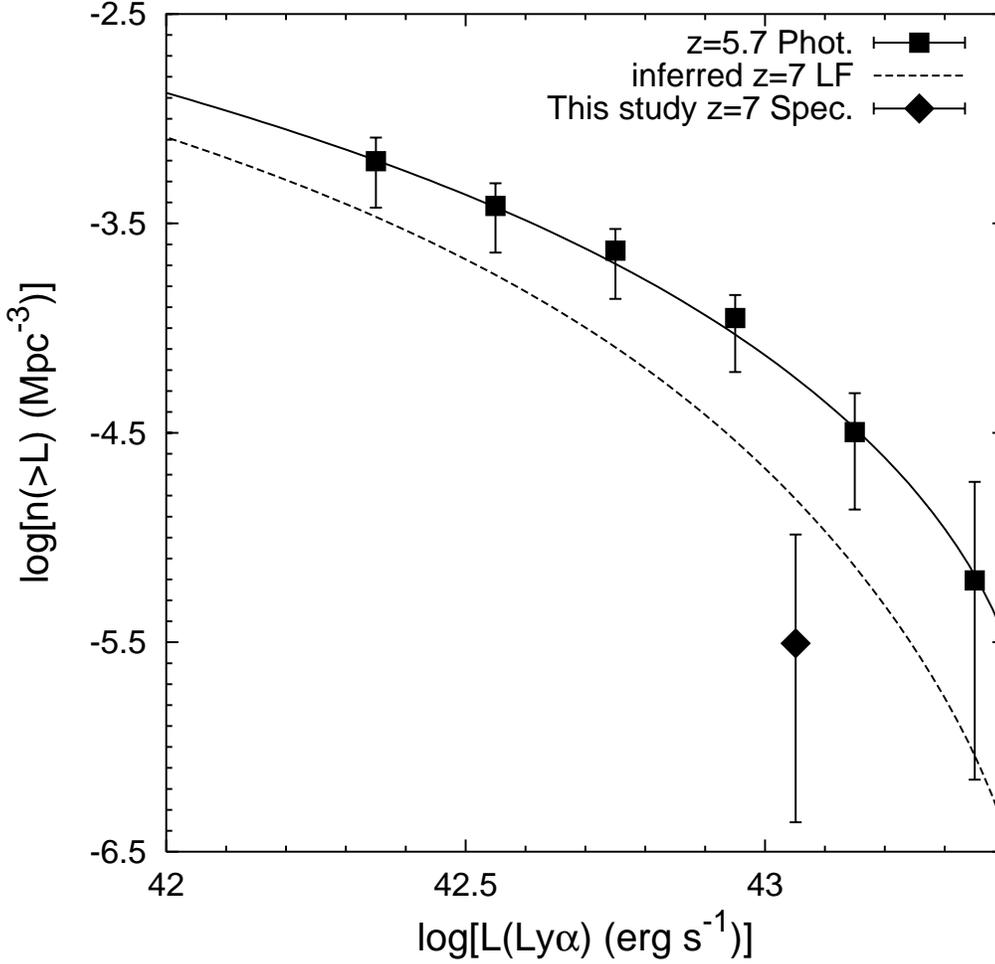}
\caption{Cumulative Ly$\alpha$ LFs of LAEs at $z=5.7$ \citep[filled squares]{Shima06} and 7 (this study; diamond).  Errors include cosmic variance and Poissonian errors for small-number statistics (See \textsection \ref{Reionization} for details).  All the data and errors are corrected for their detection completeness.   The solid curve is the best-fit $z=5.7$ Schechter LF (See \textsection \ref{Reionization}).  The dashed curve is the same Ly$\alpha$ Schechter LF but made to evolve from $z=5.7$ to $z=7$ by imposing $L^{*,\rm expect}_{z=7}\sim L^*_{z=5.7}\times 0.58$ obtained using the correlation of $\Delta M^*_{\rm UV} / \Delta z \sim 0.47$ extrapolated from the result of \citet[See their Figure 22]{Yoshida06} and assuming the same correlation can also hold for Ly$\alpha$ luminosity (See text in \textsection \ref{UVLF-Yoshida}).\label{MS_LyaLF}}
\end{figure}

The uncertainties in the number and luminosity densities at $z=5.7$--7 LAEs in the SDF in Figure \ref{MS_Kobayashi-LyaLFs} and \ref{MS_Madau_PLOT} (and likewise Figure \ref{MS_UVLF} and \ref{MS_LyaLF}) include cosmic variance and the Poissonian errors associated with small-number statistic.  To estimate the cosmic variance $\sigma_v$ at $z=5.7$--7, we adopt a bias parameter $b=3.4\pm 1.8$ derived from the sample of 515 $z=5.7$ LAEs detected in $\sim1.0$ deg$^2$ of the SXDS field \citep{Seki04,Ouch05}, which is $\sim 5 \times$ wider than SDF.  Then, applying the dark matter halo variances ($z, \sigma_{\rm DM}) = (5.7, 0.063), (6.6, 0.053)$ and (7.0, 0.044) obtained using analytic cold dark matter model \citep{Sheth99,Some04} and our SDF survey volumes to $b=\sigma_{v}/\sigma_{\rm DM}$, we calculate the geometric mean of cosmic variance at $z=5.7$--7, which is 8.4--27\%.  The maximum cosmic variance of $\sigma_v=27$\% is included in the errors in Figure \ref{MS_Kobayashi-LyaLFs}--\ref{MS_LyaLF}.  Similarly, the cosmic variance at $z=2.3$--4.5 is also calculated and included in the Figure 8.  The Poissonian errors for small-number statistic are estimated using Table 1 and 2 in \citet{Geh86}.  When the densities and errors are calculated for $z=5.7$, 6.6 and 7 LAEs in the SDF, the detection completenesses in NB816, NB921 and NB973 images are also corrected (see Figure \ref{MS_AVE_NB973_Completeness} for NB973 completeness).   

While it remains unchanged at $2.3 < z < 5.7$, the LAE number density decreases by a factor of $n_{\rm Ly\alpha}^{z=6.6}/n_{\rm Ly\alpha}^{z=5.7} \simeq 0.24$ from $z=5.7$ to 6.6 and $n_{\rm Ly\alpha}^{z=7}/n_{\rm Ly\alpha}^{z=6.6}\simeq 0.17$ from $z=6.6$ to 7.  Similarly, the LAE Ly$\alpha$ luminosity density declines by factors of $\rho_{\rm Ly\alpha}^{z=6.6}/\rho_{\rm Ly\alpha}^{z=5.7} \simeq 0.21$ and $\rho_{\rm Ly\alpha}^{z=7}/\rho_{\rm Ly\alpha}^{z=6.6} \simeq 0.15$.  If we assume that the LAE population does not evolve from $z=7$ to 5.7, this density deficit might reflect an increase in neutral IGM hydrogen with redshifts.  

However, the density decline might also possibly be ascribed to the evolution of LAE population.  If the number of LAEs having luminosities fainter than our SDF detection limits drastically increases from $z=5.7$ to 7, this can certainly affect our estimations of $n_{\rm Ly\alpha}$ and $\rho_{\rm Ly\alpha}$.  Hence, the Ly$\alpha$ LF alone cannot resolve this degeneracy between the reionization and galaxy evolution effects.

To cope with this matter, the rest frame UV continuum luminosity function (UVLF) of LAEs can be used to extrtact the galaxy evolution effect alone since it is not suppressed by neutral hydrogen.  \citet{kashik06} have compared the UVLF (rest frame $\sim1255$\AA~at $z=6.6$ and $\sim1350$\AA~at $z=5.7$) of LAEs in SDF and other field also imaged by Suprime-Cam and found that it does not significantly change from $z=5.7$ to 6.6.  This suggests that the density deficit between $z=5.7$ and 6.6 are not mainly caused by galaxy evolution.  Thus, \citet{kashik06} concluded that the reionization might have ended at around $5.7<z<6.6$ and it supports the results of $z\sim6$ quasars and GRB \citep{Fan06,Totani06}.  If this is also the case for $z=6.6$--7 LAEs, the further decline of LAE density implies increase in nuetral hydrogen that attenuates Ly$\alpha$ photons and supports \citet{kashik06}'s result.  

%\subsection{The Possibility of the Galaxy Evolution at $z=6.6$--$7$ \label{GalEvol}}
\subsection{Can the Ly$\alpha$ LF evolution be explained only by galaxy evolution? \label{GalEvol}}
We do not know whether the LAEs themselves evolve at $z=6.6$--7.  If the galaxy evolution occurs at $z=6.6$--7, the further decline of LAE density at these epochs reflects it in addition to reionization.  Hence, in this section, we investigate the possibilities of the LAE evolution from $z=6.6$ to 7 using three independent methods: (1) Comparison of the UVLFs of $z=5.7$ and 6.6 LAEs with that of $z=7$ LAEs derived from our spectroscopic data of IOK-1, (2) Estimation from the UVLF evolution of LBGs and (3) Application of an LAE evolution model constructed by \citet{ktn07} based on a hierarchical clustering galaxy formation model \citep{ny04} to predict the expected change of Ly$\alpha$ LF from $z=7$ to 5.7 due to galaxy evolution alone.  

\subsubsection{Implications from UVLF of $z=7$ LAEs \label{subsubSec_UVLF}}
First, we roughly estimate UVLF of $z=7$ LAEs to see if there is any possible galaxy evolution from $z=6.6$.  We calculate absolute UV magnitude $M_{\rm UV,1230}$ at the rest frame 1230\AA~for IOK-1 from the UV continuum flux $F({\rm UV})$ obtained in \textsection \ref{IOK-1Spec} using equation \ref{F_UV} and $F(\rm Ly\alpha)$ measured in the spectrum of IOK-1.  That is,    
\begin{eqnarray}
M_{\rm UV,1230} &=& m_{\rm UV,1230}-DM+2.5\log(1+z) \nonumber \\ 
&=& -2.5\log\left[\frac{\lambda^2}{c\Delta \lambda}F({\rm UV})\right]-48.6-DM+2.5\log(1+z) 
\label{AbsMag_UV_SPEC}
\end{eqnarray}
where $m_{\rm UV,1230}$ is UV apparent magnitude, $\lambda=1230(1+z)${\rm\AA~}and $\Delta \lambda$ is the wavelength range in which the UV continuum is covered by NB973 passband, $\Delta \lambda=9855{\rm \AA}-(1+z)1216{\rm\AA}$.  Also, $DM$ and $c$ are a distance modulus and the speed of light.  
%For comparison, we also derive $M_{UV,1230}$ of IOK-1 and -2 using the photometric model presented in \textsection \ref{PHOT_LAE_Model}.  The equation \ref{F-PHOT-UV} is used to convert the fluxe densities, $f^{NB973}_{\nu}$, of IOK-1 and -2 (corresponding to their NB973 total magnitudes) into their UV absolute magnitudes $M_{UV,1230}$ as follows.  
%\begin{equation}
%M_{UV,1230} = -2.5\log(0.31f^{NB973}_{\nu})-48.6-DM+2.5\log(1+z)
%\label{AbsMag_UV_PHOT}
%\end{equation}
Figure \ref{MS_UVLF} shows the UVLF of $z=7$ LAE derived here together with those of $z=5.7$ and 6.6 LAEs.  We ignore a subtle difference in the rest frame UV wavelengths (rest frame $\sim1230$\AA~at $z=7.0$, $\sim1255$\AA~at $z=6.6$ and $\sim1350$\AA~at $z=5.7$) assuming that the LAEs have flat UV continua.  Also, the detection completeness of NB973 image is corrected using Figure \ref{MS_AVE_NB973_Completeness}.  The UVLF implies that there is no galaxy evolution from $z=7$ to 6.6, and the density deficits of $n_{\rm Ly\alpha}$ and $\rho_{\rm Ly\alpha}$ between these epochs might be attributed mainly to the reionization.

\subsubsection{Estimation from the UVLF Evolution of LBGs \label{UVLF-Yoshida}}
Even though the $z=7$ UVLF derived from the SDF data suggests that LAEs do not evolve from $z=7$ to 6.6, it suffers small statistics due to the relatively shallower detection limit in NB973 (equivalent to $L(\rm Ly\alpha) \geq 1.0\times 10^{34}$ erg s$^{-1}$).  Therefore, we discuss the possibilities of the LAE galaxy evolution at $z=5.7$--7 using inferences from other independent methods, by which we try to obtain some helpful insights.

One possible way to estimate the LAE galaxy evolution at $z=5.7$--7 is the inference from the evolution of UVLF of high-$z$ LBGs, assuming LAEs and LBGs share a similar evolutionary history.  We use the UVLF data from the recent observational studies about $z\sim 4$--8 LBGs conducted by \citet{Yoshida06}, \citet{Bouw06} and \citet{06BI}.  Their surveys, when combined together, form the deepest and the widest imaging data with the samples of the largest numbers in all LBG surveys.  Interestingly, \citet{Yoshida06} combined their data of $z\sim4$ and 5 LBGs with those from lower-$z$ LBG surveys and $z\sim6$ LBG ($i$-dropout) study by \citet{Bouw06} and found clear evolution of the UVLF from $z\sim 6$ to 0, in which only the characteristic magnitude, $M_{\rm UV}^*$, changes significantly and almost linearly with redshift while the normalization factor, $\phi^*$, and the faint end slope, $\alpha$, tend to remain constant as seen in Figure 22 of \citet{Yoshida06}.  This trend of $M_{\rm UV}^*$, $\phi^*$ and $\alpha$ vs. $z$ continues up to $z\sim 7.4$ when we add $M_{\rm UV}^*=-19.5\pm0.6$ mag, $\phi=0.00202^{+0.00086}_{-0.00076}$ Mpc$^{-3}$ and $\alpha=-1.73$ (or $M_{\rm UV}^*=-18.75\pm0.6$ mag, $\phi=0.00218$ Mpc$^{-3}$ and $\alpha=-1.73$) of the first (or second) LBG UVLF at $z\sim7.4$ derived by \citet{06BI}.   

We estimate the change in $M_{\rm UV}^*$ between $z=5.7$ and $z=7$ from the $z$-dependence of $M_{\rm UV}^*$ at $z\sim 4$, 5, 6 and 7.4, assuming the correlation is linear with the slope of $\Delta M^*_{\rm UV}/\Delta z\simeq 0.47$ mag.  As a result, $M^*_{\rm UV}$ is expected to become fainter by 0.6 mag, which corresponds to the luminosity of $L^{*,\rm expect}_{z=7} \simeq L^*_{z=5.7}\times 10^{-0.4\times0.6}\simeq L^*_{z=5.7}\times 0.58$.  Here, the relation between the equations \ref{Scheter-LF} and the Schechter LF in absolute magnitude form,
\begin{equation}
\phi (M)dM=\frac{2}{5}\phi^*(\ln10)\left[10^{\frac{2}{5}(M^*-M)}\right]^{\alpha +1}\exp\left[-10^{\frac{2}{5}(M^*-M)}\right]dM
\label{AbsMag_SchechterLF}
\end{equation}
is used to convert $M^*_{\rm UV}$ to $L^*$.
To infer the deficit by which Ly$\alpha$ LF of LAEs decreases from $z=5.7$ to 7 due to thier evolution alone, we now roughly assume that this UVLF evolution of the LBGs can be also applied to that of LAEs at $z=5.7$--7 and Ly$\alpha$ line luminosities of LAEs are simply proportional to their UV continuum luminosities as Figure 15 in \citet{Tani05} suggests.  Based on this idea, we change $L^*_{z=5.7}$ of our best-fit Schechter Ly$\alpha$ LF at $z=5.7$ in exactly the same way (i.e., $\log(\phi^* [$Mpc$^{-3}])=-3.44^{+0.20}_{-0.16}$ and $\alpha=-1.5$ as in \textsection \ref{Reionization} but $\log(L^*_{z=5.7} [$erg s$^{-1}])=43.04^{+0.12}_{-0.14}+\log0.58$ this time) to obtain $z=7$ Ly$\alpha$ LF.  This result is compared in Figure \ref{MS_LyaLF} with actual observation data of IOK-1.  

The inferred Ly$\alpha$ LF at $z=7$ does not really agree with one calculated from the spectrum of IOK-1.  Our density deficit between $z=5.7$ and 7 LAEs cannot be explained by only the galaxy evolution factor estimated here.  The integrations of the inferred Ly$\alpha$ LF using equation \ref{integ-Scheter-LF} and \ref{integ-L*Scheter-LF} down to $\log L(\rm Ly\alpha)=43.05$, which is IOK-1's Ly$\alpha$ line luminosity, yield $n^{{\rm expect},z=7}_{\rm Ly\alpha} \simeq 1.5\times 10^{-5}$ Mpc$^{-3}$ and $\rho^{{\rm expect},z=7}_{\rm Ly\alpha} \simeq 2.3\times 10^{38}$ erg s$^{-1}$Mpc$^{-3}$, respectively.  Our LAE number and Ly$\alpha$ line luminosity densities at $z=7$ are $n^{z=7}_{\rm Ly\alpha} \simeq (3.6_{-2.8}^{+7.3})\times 10^{-6}$ Mpc$^{-3}$ and $\rho^{z=7}_{\rm Ly\alpha} \simeq (4.1_{-3.1}^{+8.2})\times 10^{37}$ erg s$^{-1}$Mpc$^{-3}$ based on IOK-1 data alone, respectively.  Therefore, the density deficits of $n^{z=7}_{\rm Ly\alpha}/n^{{\rm expect},z=7}_{\rm Ly\alpha} \simeq 0.24_{-0.19}^{+0.49}$ and $\rho^{z=7}_{\rm Ly\alpha}/\rho^{{\rm expect},z=7}_{\rm Ly\alpha} \simeq 0.18_{-0.13}^{+0.36}$ might be due to the attenuation of Ly$\alpha$ photons by neutral IGM having existed during the reionization.  

In order for the inferred Ly$\alpha$ LF at $z=7$ to have the same number density as observed Ly$\alpha$ LF at $z=7$ (i.e., $n^{{\rm expect},z=7}_{{\rm Ly\alpha}}=n^{z=7}_{\rm Ly\alpha}$), we have to change the characteristic luminosity $L^{*,\rm expect}_{z=7}$ by a factor of $\times 0.65_{-0.18}^{+0.24}$.  This factor might reflect the deficit by which Ly$\alpha$ LF of LAEs decreases from $z=5.7$ to 7 due to the attenuation of the Ly$\alpha$ lines of LAEs by the increasing neutral IGM during the reionization beyond $z\sim6$.  We can refer to such a deficit factor due to the neutral IGM attenuation as IGM transmission to Ly$\alpha$ photons, $T_{\rm Ly\alpha}^{\rm IGM}$.  In the case of our discussion so far, this can be regarded as the ratio of the Ly$\alpha$ line luminosities of LAEs in the environments with some neutral IGM fraction $x_{\rm HI}$ still remaining and with no neutral IGM ({\rm i.e.}, $x_{\rm HI}=0$), $T_{\rm Ly\alpha}^{\rm IGM} = L^{x_{\rm HI}}({\rm Ly\alpha})/L^{x_{\rm HI}=0}({\rm Ly\alpha})$.  Once we know $T_{\rm Ly\alpha}^{\rm IGM}$, the neutral IGM fraction at $z=7$, $x_{\rm HI}^{z=7}$, can be estimated.  However, the calculation of $x_{\rm HI}$ from $T_{\rm Ly\alpha}^{\rm IGM}$ is not a simple issue and is dependent on theoretical models.  We will discuss it in \textsection \ref{neutral-IGM}. 

Similarly, the $z$-dependence of $M_{\rm UV}^*$, $\Delta M^*_{\rm UV}/\Delta z\simeq 0.47$ mag, predicts $\Delta M^*_{\rm UV}\simeq 0.42$ for $\Delta z=6.6-5.7$ and thus $L^{*,{\rm expect}}_{z=6.6} \simeq L^*_{z=5.7}\times 10^{-0.4\times0.42}\simeq L^*_{z=5.7}\times 0.68$ due to LAE galaxy evolution from $z=6.6$ to 5.7.  However, \citet{kashik06} found that the Ly$\alpha$ LF declines in such a way that $L^*_{z=6.6} \sim L^*_{z=5.7}\times (0.4$--0.6) from $z=5.7$ to 6.6, regarding their photometric and spectroscopic LFs as the upper and lower limits of $z=6.6$ Ly$\alpha$ LF, respectively.  Hence, the attenuation of Ly$\alpha$ photons by the neutral IGM at $z=6.6$ is $T_{\rm Ly\alpha}^{\rm IGM}=L^*_{z=6.6}/L^{*,{\rm expect}}_{z=6.6} \simeq 0.59$--0.88. 
   
The decrease in Ly$\alpha$ LF from $z=5.7$ to 6.6 and 7 cannot be explained only by the evolution of LAEs inferred from that of LBGs.  This result implies that the remaining deficits could come from the attenuation of Ly$\alpha$ lines by neutral IGM.  If this is the case, Ly$\alpha$ line tends to be more attenuated at higher redshift as $T_{\rm Ly\alpha}^{\rm IGM}\simeq 0.59$--0.88 at $z=6.6$ and $0.65_{-0.18}^{+0.24}$ at $z=7$, implying that the neutral IGM fraction, $x_{\rm HI}$ increases with redshift beyond $z\sim6$ as derived in \textsection \ref{neutral-IGM}.  However, note that this result is based on the assumption that LAEs evolve in the same way as LBGs do.  This might not be necessarily true.  Although the LAEs are believed to be closely related to LBGs and many of candidate LBGs at high redshift have been identified as LAEs by spectroscopy, the link between these two populations has not been clearly understood yet and they might have followed different evolutionary histories.     
                
\subsubsection{Application of A Galaxy Evolution Model \label{Model_GalEvol}} 
In the previous sections, we tried to estimate the intrinsic evolution
of Ly$\alpha$ LF of LAEs from UVLF evolution
of LAEs and LBGs, with an implicit
assumption that the evolutions of Ly$\alpha$ and UV luminosities are
similar to each other.  However, this assumption may not be true in
reality, and hence our argument will be strengthened if we can show that
these are indeed similar in a realistic theoretical model of LAEs.

For this purpose, we use a recent model for LAE evolution constructed by \citet[hereafter K07]{ktn07}.  This model is an extension of
one of the latest hierarchical clustering models of galaxy formation \citep{ny04}, in which the merger histories of dark matter
haloes are modeled based on structure formation theory and star
formation processes in dark haloes are calculated to predict the
photometric properties of galaxies. This model can reproduce most of the
observations for photometric, kinematic, structural, and chemical
properties of local galaxies, as well as high-$z$ LBGs \citep{kashikawa06}.
K07 extended this model without changing the original model parameters,
but introducing new modeling only for the escape fraction of
Ly$\alpha$ photons ($f_{\rm esc}^{\rm Ly \alpha}$) from galaxies based
on physical considerations. Specifically, the dust extinction of
Ly$\alpha$ photons and effect of galaxy-scale outflow are newly taken
into account.  This is the first model for LAEs based on a
hierarchical galaxy formation model in which $f_{\rm esc}^{\rm Ly
\alpha}$ is not a universal constant but depends on physical conditions
of galaxies.  This model can reproduce the observed Ly$\alpha$ LF of
LAEs in $z \sim $ 3--6, and predicts that galaxies under strong galaxy
scale outflow with $f_{\rm esc}^{\rm Ly \alpha} \sim 1$ are dominant in
the bright-end of Ly$\alpha$ LFs, which is also consistent with
observations.  It should be noted here that $f_{\rm esc}^{\rm Ly\alpha}$
in the K07 model could vary from galaxy to galaxy and may evolve
within a galaxy, and hence even if the Ly$\alpha$ photon production rate
is proportional to star formation rate, the evolutions of Ly$\alpha$ and
UV LFs could be different from each other.

The K07 predictions of the Ly$\alpha$ and UV LFs of LAEs at $z=5.7,~6.6$
and $7$ assuming $T_{\rm Ly\alpha}^{\rm IGM} = 1$ are presented in
Figure \ref{MS_Kobayashi-LyaLFs} and \ref{MS_UVLF}, respectively.  The evolutions of number density and
Ly$\alpha$ luminosity density of LAEs with a threshold Ly$\alpha$
luminosity predicted by this model are shown Figure \ref{MS_Madau_PLOT}.  As demonstrated
in K07, the deficit of the observed LAEs compared with the model
prediction of Ly$\alpha$ LF is clear at $z \gtrsim 6$ as seen in Figure \ref{MS_Kobayashi-LyaLFs} while this model precisely reproduces the observed evolution at $z \sim$3--6.  On the
other hand, the degree of evolution of UVLF of LAEs predicted by the
model is similar to that observed in the same redshift range.

The fact that the model prediction
is consistent with the UVLF evolution but not
with the Ly$\alpha$ LF evolution then implies that the evolution of the observed Ly$\alpha$ LF at $z \lesssim 6$ could be caused by the IGM absorption.
The discrepancy can be resolved if we adopt a simple prescription of
luminosity-independent IGM transmission:
$T_\mathrm{Ly\alpha}^\mathrm{IGM} = 0.62$--0.78 at $z=6.6$ and
$T_\mathrm{Ly\alpha}^\mathrm{IGM} = 0.40$--0.64 at $z=7$.

\subsection{Implications for Reionization \label{neutral-IGM}}
In the previous section we have shown that the evolution
of Ly$\alpha$ LF at $z \gtrsim 6$ could be likely a result of Ly$\alpha$
photon absorption by neutral IGM, implying a significant evolution of
the IGM neutral fraction beyond $z \gtrsim 6$.  In order to obtain some
quantitative implications for reionization, however, we must
translate the estimates of $T_\mathrm{Ly\alpha}^\mathrm{IGM}$ obtained
in the previous section into the IGM neutral fraction, $x_{\rm
HI}$. This procedure is not straightforward because this translation is
generally model dependent (e.g., Santos 2004; Dijkstra et al. 2007).

Here, we apply the dynamic model with a reasonable velocity shift of
Ly$\alpha$ line by $360~\mathrm{km~s^{-1}}$ redward of the systemic
velocity \citep{San04}.  The attenuation factor of Ly$\alpha$ luminosity
is given as a function of $x_{\rm HI}$, and the reason for the choice of
this model is that this model predicts no attenuation when $x_{\rm HI} =
0$.  Note that some other models of Santos (2004) predict a significant
attenuation even in the case of $x_{\rm HI}=0$, due to the neutral gas
associated with the host haloes of LAEs. Choosing this particular model
then means that we ascribe the evolution of the Ly$\alpha$
LF at $z \gtrsim 6$ only to the absorption by pure IGM. We consider that
this is a reasonable assumption, since observations indicate that the
escape fraction of Ly$\alpha$ photons is about unity at least for LAEs
at $z \sim 3$ \citep{Gawiser06}. If LAEs at $z \sim 7$ are a similar
population to the low-$z$ LAEs, we do not expect significant absorption
by neutral gas physically associated to LAEs. On the other hand, it
should also be kept in mind that if $z\sim 7$ LAEs are surrounded by a
significant amount of nearby neutral gas that is not present for low-$z$
LAEs, the estimate of $x_{\rm HI}$ as an average of IGM in the universe
could become lower than those derived here.

In section \ref{UVLF-Yoshida}, we obtained $T_\mathrm{Ly\alpha}^\mathrm{IGM}=0.59$--0.88 and $0.65_{-0.18}^{+0.24}$ at $z=6.6$ and 7.0, respectively.  Application of \citet{San04} model yields the neutral fractions of $x_{\rm HI}^{z=6.6}\sim 0.12$--0.42 and $x_{\rm HI}^{z=7}\sim 0.12$--0.54.
If we use the K07 model in section \ref{Model_GalEvol} to estimate $x_{\rm HI}$, we find $x^{z=6.6}_\mathrm{HI} \sim 0.24$--0.36 from $T_\mathrm{Ly\alpha}^\mathrm{IGM}=0.62$--0.78 at $z=6.6$, and $x^{z=7}_\mathrm{HI} \sim 0.32$--0.64 from $T_\mathrm{Ly\alpha}^\mathrm{IGM} =0.40$--0.64 at $z =7$.  The neutral fraction $x_{\rm HI}$ at $z=6.6$ and 7 estimated from two independent methods are consistent with each other and tends to increase with redshift at $z>6$.  These series of $x_{\rm HI}$ values at $z=6.6$ and 7, combined with $x_{\rm HI}^{z\sim6.2} \sim 0.01$--0.04 and $x_{\rm HI}^{z\sim6.3} < 0.17$--0.6 derived from quasar GP tests and GRB spectral analysis \citep{Fan06,Totani06}, supports the picture that the reionization completed at $z\sim6$, beyond which it was still in progress with larger neutral fraction of IGM hydrogen, which evolved with redshift.  The neutral IGM fractions obtained by independent methods are summerized in Table \ref{Neutral_Frction}.  

However, our constraint suggests that the neutral IGM persists at the $\sim50$\% level as late as $z=7$ and would contradict the {\it WMAP} conclusion that the reionization epoch was $z = 10.9^{+2.7}_{-2.3}$ at the $>95$\% confidence level.  Our results could be reconciled with {\it WMAP} only if there is a statistical fluke (one time in 20, a 95\% confidence range is wrong) or the reionization happened twice (e.g., Cen 2003) so that a lot of the observed electron scattering happens at $z \gg 7$, and then the universe becomes partially neutral again, allowing us to observe neutral gas at $z=7$.    

Finally, we again emphasize that these quantitative results
are model-dependent and should be interpreted with caution.
However, the decrease in the Ly$\alpha$ LF of LAEs beyond
$z \sim 6$ is more significant than expected from
UVLF evolution or a theoretical model, and hence
the physical status of IGM might be changing at $z \gtrsim 6$.

\section{Summary and Conclusion}
 We have conducted a narrowband NB973 survey of $z=7$ LAEs, established color criteria to select out $z=7$ LAEs, and found two candidates down to $L(\rm Ly\alpha) \geq 1.0 \times 10^{43}$ erg s$^{-1}$ (5$\sigma$). By follow-up spectroscopy, the brighter of the two was indentified as a $z=6.96$ LAE while we can confirm neither Ly$\alpha$ emission nor any other features in the spectrum of the other candidate despite the sufficiently long integration time.  

The number and Ly$\alpha$ luminosity densities at $z=7$ obtained by this study were compared to those at $z=5.7$ and 6.6 derived from the latest samples obtained by the SDF surveys \citep{Shima06,kashik06} down to our detection limit, and clear evolution of density deficits with increasing redshifts was observed such that: $n_{\rm Ly\alpha}^{z=6.6}/n_{\rm Ly\alpha}^{z=5.7} \simeq 0.24$ and $n_{\rm Ly\alpha}^{z=7}/n_{\rm Ly\alpha}^{z=6.6}\simeq 0.17 $; $\rho_{\rm Ly\alpha}^{z=6.6}/\rho_{\rm Ly\alpha}^{z=5.7} \simeq 0.21$ and $\rho_{\rm Ly\alpha}^{z=7}/\rho_{\rm Ly\alpha}^{z=6.6} \simeq 0.15$.  If we assume that the LAE population does not evolve from $z=7$ to 5.7, this series of density deficits could reflect an increase in neutral IGM hydrogen with redshifts beyond $z\sim 6$.

To see if LAE evolves from $z=7$ to 6.6, we also compared UVLF of $z=7$ LAE with those of $z=5.7$ and 6.6 LAEs derived from the SDF LAE surveys.  No decrease in the number density of UVLF was observed from $z=5.7$ through 6.6 to 7.  Since the UV photons are not attenuated by neutral IGM and the UVLF is only sensitive to galaxy evolution, our result suggests that the deficits in $n_{\rm Ly\alpha}$ and $\rho_{\rm Ly\alpha}$ might reflect the cosmic reionization and the LAE population does not significantly evolve at $z=5.7$--7.    

However, the UVLF at $z=7$ suffers from small statistics at this time and the interpretation is not robust.  Hence, the amount by which the LAE evolution affects the density deficits among $z=7$, 6.6 and 5.7 were investigated by the inference from UVLFs of $z<8$ LBGs \citep{Yoshida06,06BI,Bouw06} based on the assumption that LAEs would have evolved in the same way as LBGs and Ly$\alpha$ line luminosities of LAEs are proportional to their UV continuum luminosities.  Even after the galaxy evolution was taken into account, there still remained some density deficits among these epochs.  If we attribute the deficits to the attenuation of Ly$\alpha$ photons by the neutral IGMs, the neutral fractions of the Universe at $z=6.6$ and 7 are estimated to be 0.12--0.42 and 0.12--0.54, respectively.  This result, combined with neutral fractions derived from $z\sim 6$ quasars and a $z\sim6.3$ GRB, supports the completion of the reionization at $z\sim 6$ and the possible evolution of neutral IGM beyond this redshift.

Again, this result is based on the assumption that LAEs would have evolved in the same way as LBGs, which might not be necessarily true.  Therefore, we furthermore used a LAE evolution model (K07 model) constructed from hierarchical clustering scenario to reproduce Ly$\alpha$ LFs at $z=5.7$, 6.6 and 7 in the case of transparent IGM ($x_{\rm HI}=0$) and compared with Ly$\alpha$ LFs obtained by the latest SDF surveys \citep{Shima06,kashik06,06IOK}.  The observed data at $z=6.6$ and 7 showed smaller number and luminosity densities than those predicted by the model, suggesting that there still remains the possibility of the incomplete reionization at those epochs.  The neutral fractions at $z=6.6$ and 7 estimated from the decline of the LFs by the reionization factors alone after the galaxy evolution effects had been corrected are $x_{\rm HI}^{z=6.6}\sim 0.24$--0.36 and $x_{\rm HI}^{z=7}\sim 0.32$--0.64, respectively, also consistent with quasar and GRB results.  

The results regarding $z=7$ LAE presented here are based on relatively shallow depth of NB973 imaging, small sample statistics, and only the SDF and optical imaging data.  From these data alone, the trend of the density deficit between $z=5.7$ and 7 in fainter LAE populations and in other places in the Universe, changes in physical properties of LAEs associated with their evolution between these epochs, and typical spectroscopic properties of $z=7$ LAEs such as the direct detection of the attenuation of Ly$\alpha$ line cannot be inferred.  Also, the calculation of the UV continuum flux and thus UVLF is dependent on relatively rough estimation without infrared data.  Deeper NB973 imaging of the SDF as well as other fields for which infrared images are available and follow-up spectroscopy of newly detected LAE candidates will provide the answers and more precise results in future studies.

%Combining all the suggested surveys and observations, we expect to investigate the variation of the reionization and galaxy evolution status over the cosmic time ($z=$5.7--7) and space (SDF, SXDF, and GOODS fields), which is very important to unvail the observationally highly unexplored epochs in the early Universe.  

%% Included in this acknowledgments section are examples of the
%% AASTeX hypertext markup commands. Use \url without the optional [HREF]
%% argument when you want to print the url directly in the text. Otherwise$z=7$ LAE,
%% use either \url or \anchor, with the HREF as the first argument and the
%% text to be printed in the second.

\acknowledgments

We greatly appreciate the technology and engineers of Asahi Spectra Co., Ltd.~for developing the NB973 filter that led us to the discovery of the $z=6.96$ LAE.  We are deeply grateful to the staff at the Subaru Telescope for their kind supports to make our observations successful.  We express the gratitude to the SDF team for obtaining and providing us with invaluable imaging data.  K.O. acknowledges the fellowship support from the Japan Society for the Promotion of Science and the Special Postdoctoral Researchers Program at RIKEN.
%We also thank the anonymous referee for useful comments that significantly improved this paper.



\clearpage

%% Tables should be submitted one per page, so put a \clearpage before

%% Two options are available to the author for producing tables:  the
%% deluxetable environment provided by the AASTeX package or the LaTeX
%% table environment.  Use of deluxetable is preferred.
%%

%% Three table samples follow, two marked up in the deluxetable environment,
%% one marked up as a LaTeX table.

%% If you use the table environment, please indicate horizontal rules using
%% Do not put multiple tabular environments within a single table.
%% The optional \label should appear inside the \caption command.

\clearpage

\begin{table}
\begin{center}
\small
\caption{Photometric properties of candidate Ly$\alpha$ emitters and NB973-excess objects\label{Photo-property}}
\begin{tabular}{lcrrrrccc}\tableline\tableline
Object                 & Coordinate             & $i'$     & NB816 & $z'$        & NB921    & NB973 & NB973 (total) & Criteria\\\tableline
IOK-1\tablenotemark{a} & 13:23:59.8 +27:24:55.8 & $>$27.84 & $>$27.04 & $>$27.04 & $>$26.96 & 24.60 & 24.40         & (1)\\
IOK-2                  & 13:25:32.9 +27:31:44.7 & $>$27.84 & $>$27.04 & $>$27.04 & $>$26.96 & 25.51 & 24.74         & (1)\\\tableline
IOK-3\tablenotemark{b} & 13:24:10.8 +27:19:28.1 & $>$27.84 & $>$27.04 & 26.26    & 25.08    & 24.87 & 24.57         & ---\tablenotemark{c}\\ 
Obj-4\tablenotemark{d} & 13:25:09.1 +27:32:16.8 & 27.14    & 26.77        & 25.75    & 25.51    & 24.97 & 24.85         & (3)\\
Obj-5\tablenotemark{d} & 13:23:45.8 +27:32:51.4 & $>$27.84 & $>$27.04 & 25.76    & 25.44    & 25.10 & 24.74         & (3)\\\tableline
\end{tabular}
\tablenotetext{}{NOTE: Units of coordinate are hours: minutes: seconds (right ascension) and degrees: arcminutes: arcseconds (declination) using J2000.0 equinox.  $i'$, NB816, $z'$, NB921 and NB973 are all $2''$ aperture magnitudes while NB973 (total) is a total magnitude.  Magnitudes are replaced by their $2\sigma$ limits if they are fainter than the limits.  Color criteria are those used to select out candidate LAEs and classify $z'-$ NB973 excess objects (See \textsection \ref{COLOR_CRITERIA} and \ref{NB973_EXCESS_OBS}).} 
\tablenotetext{a}{IOK-1 was proven to be a $z=6.96$ LAE spectroscopically by \citet{06IOK}.}
\tablenotetext{b}{IOK-3 was found to be a $z=6.6$ LAE by independent spectroscopy by \citet{kashik06}}
\tablenotetext{c}{IOK-3 satisfies color criteria (1) and (2) simultaneously except for NB921 $<3\sigma$.}
\tablenotetext{d}{These objects show flux excess of $1>z'-$ NB973 $>3\sigma$ but could be either of $z=6.2$--6.4 LAEs, low-$z$ ellipticals or late-type stars (See \textsection \ref{NB973_EXCESS_OBS}).}
%% Any table notes must follow the \end{tabular} command.
\end{center}
\end{table}

\clearpage

%\begin{table}
%\begin{center}
%\caption{Star formation properties of $z=7$ LAEs estimated from NB973 photometric model\label{Photo-F-L-SFR}}
%\begin{tabular}{lccccc}\tableline\tableline
%Object & $F^{phot}(Ly\alpha)$               & $L^{phot}(Ly\alpha)$    & $SFR^{phot}(Ly\alpha)$  & $L^{phot}_{\nu}(UV)$ & $SFR^{phot}(UV)$\\
%       & (10$^{-17}$erg s$^{-1}$ cm$^{-2}$) & (10$^{43}$erg s$^{-1}$) & (M$_{\odot}$ yr$^{-1}$) & (10$^{29}$erg s$^{-1}$ Hz$^{-1}$) & (M$_{\odot}$ yr$^{-1}$)\\\tableline
%IOK-1  & 2.71                               & 1.52                    & 13.87                   & 1.38                 & 19.3\\
%IOK-2  & 1.98                               & 1.14                    & 10.32                   & 1.03                 & 14.4\\\tableline
%\end{tabular}
%% Any table notes must follow the \end{tabular} command.
%\end{center}
%\end{table}
%
%\clearpage
%
%\begin{table}
%\begin{center}
%\caption{Full width at half maximum size of candidate Ly$\alpha$ emitters\label{table2}}
%\begin{tabular}{lcccc}\tableline\tableline
%Object & $FWHM$ & $FWHM_{cor}$ & $FWHM_{cor}^*$ & Stellarity\\\tableline
%IOK-1  &        &  &  &   \\
%IOK-2  &        &  &  &   \\\tableline
%IOK-3  &        &  &  &    \\ 
%Obj-4  &        &  &  &   \\
%Obj-5  &        &  &  &   \\\tableline
%\end{tabular}
%\tablenotetext{*}{Deconvolved with a PSF size of $0.''78$}
%% Any table notes must follow the \end{tabular} command.
%\end{center}
%\end{table}

\clearpage

\begin{table}
\begin{center}
\caption{The number counts of $i'$-detected objects fainter than our detection limit NB973 $=24.9$\label{N_vs_i}}
\begin{tabular}{cccc}\tableline\tableline
$i'^a$       & NB973$^a$        & $\Delta m^b$ & $N(i')=N(\Delta m)$$^c$ \\\tableline
25.23        & 24.9 $(5\sigma)$ & 0.0          & --- \\
25.23--25.33 & 24.9--25.0       & 0.0--0.1     & 2297\\
25.33--25.83 & 25.0--25.5       & 0.1--0.6     & 13589\\ 
25.83--26.33 & 25.5--26.0       & 0.6--1.1     & 16012\\
26.33--26.83 & 26.0--26.5       & 1.1--1.6     & 17367\\\tableline
\end{tabular}
\tablenotetext{}{NOTE: Units of the first to third columns are all AB magnitudes with $2''$ aperture.}
\tablenotetext{a}{$i'$ band magnitudes were converted into NB973, using $< i'-$ NB973 $>=0.33$.}
\tablenotetext{b}{The NB973 magnitude increments required for objects to become brighter than our detection limit NB973 $(5\sigma)=24.9$.  That is, $\Delta m =$ NB973 $- 24.9$ or $\Delta m =i'- 25.23$.}
\tablenotetext{c}{The number counts of $i'$-detected objects.}
%% Any table notes must follow the \end{tabular} command.
\end{center}
\end{table}

\clearpage

\begin{table}
\begin{center}
\caption{The number counts of objects in the NB973 and $i'$ images of the SDF\label{N_vs_NB973_or_i}}
\begin{tabular}{cccr}\tableline\tableline
NB973 $=i'-0.33$$^a$ & $i'$$^b$     & $N($NB973)$^c$ & $N(i')$$^d$ \\\tableline
22.5--23.0           & 22.83--23.33 & 2599         &  2588 \\
23.0--23.5           & 23.33--23.83 & 3808         &  3953 \\ 
23.5--24.0           & 23.83--24.33 & 5230         &  5697 \\
24.0--24.5           & 24.33--24.83 & 7093         &  8121 \\
24.5--25.0           & 24.83--25.33 & 8800         & 10660 \\\tableline
25.0--25.5           & 25.33--25.83 & ---          & 13589 \\
25.5--26.0           & 25.83--26.33 & ---          & 16012 \\
26.0--26.5           & 26.33--26.83 & ---          & 17367 \\\tableline
\end{tabular}
\tablenotetext{}{NOTE: Units of the first and second columns are total magnitudes.}
\tablenotetext{a}{$i'$ band magnitudes in the second column were converted into NB973 magnitudes in the first column, using $< i'-$ NB973 $>=0.33$.}
\tablenotetext{b}{$i'$ magnitudes brighter then $5\sigma$ limiting magnitude of the SDF $i'$ band image $i'(5\sigma)=26.85$.}
\tablenotetext{c}{Number counts of objects detected in the SDF NB973 image down to our detection limit NB973 $(5\sigma)=24.9$.}
\tablenotetext{d}{Number counts of objects detected in the SDF $i'$ image down to its limiting magnitude $i'(5\sigma)=26.85$.}

%% Any table notes must follow the \end{tabular} command.
\end{center}
\end{table}

\clearpage

\begin{table}
\begin{center}
\caption{The number counts of $i'$-detected variables against $i'$ brightness increment\label{N_Variables}}
\begin{tabular}{ccccccc}\tableline\tableline
$\Delta i'^a$ & \multicolumn{2}{c}{Number of variables $N_v(\Delta i')$} & \multicolumn{2}{c}{$P(\Delta i')^b=N_v(\Delta i')/N_{obj}$} &\multicolumn{2}{c}{$P(\Delta i')\times N(\Delta m)$}\\
(AB mag)     & 2003--2005 & 2001--2002 & 2003--2005 & 2001--2002 & 2003--2005 & 2001--2002\\\tableline
0.0          & ---        & ---        & ---                & ---                & ---     & ---\\
0.0--0.1     & 250        & 409        & $3.3\times10^{-3}$ & $6.8\times10^{-3}$ & 7.6$^c$ & 15.6$^c$\\
0.1--0.6     &  52        &  37        & $6.9\times10^{-4}$ & $6.2\times10^{-4}$ & 9.4     & 8.4\\ 
0.6--1.1     &   1        &   2        & $1.3\times10^{-5}$ & $3.3\times10^{-5}$ & 0.21    & 0.53\\
1.1--1.6     &   1        &   1        & $1.3\times10^{-5}$ & $1.7\times10^{-5}$ & 0.23    & 0.30\\\tableline
\multicolumn{5}{c}{Number of variables that became brighter than NB973 $=24.9$}  & 9.8$^d$ & 9.2$^d$\\\tableline
\end{tabular}
\tablenotetext{}{NOTE: The number of variables brighter than our detection limit in the stacked $i'$ SDF image, $i'=0.33+$ NB973 $(5\sigma)=25.23$, were counted.} 
\tablenotetext{a}{Increase in $i'$ magnitude over the two epochs, which is binned to match the third column ($\Delta m$) of Table 3.}
\tablenotetext{b}{Probability of finding a variable with a brightness increase of $\Delta i'=\Delta m$ in SDF down to our detection limit.}
\tablenotetext{c}{We ignore these values since the $\Delta m =0$--0.1 cannot be distinguished from photometric errors.}
\tablenotetext{d}{These figures were obtained by $\Sigma_{\Delta m}P(\Delta i')\times N(\Delta m)=\Sigma_{\Delta i}P(\Delta i')\times N(\Delta m)$.}
%% Any table notes must follow the \end{tabular} command.
\end{center}
\end{table}

\clearpage

\begin{table}
\begin{center}
\caption{Status of the follow-up spectroscopy\label{Spec-Status}}
\begin{tabular}{lrcrc}\tableline\tableline
Object    & date             & seeing             & exposure$^a$ & FOCAS Mask\\
          & (HST)            & ($''$)             & (seconds)      & \\\tableline
IOK-1$^b$ & 14, 15 May 2005  & 0.5--0.7, 0.9--1.0 & 10800         & MOS-1\\
          & 1 June 2005      & 0.6--0.8           & 3600          &      \\
          & 24 April 2006    & 0.9--1.5           & 16200         &      \\
IOK-2$^c$ & 14, 15 May 2005  & 0.5--0.7, 0.9--1.0 & 3600          & MOS-4\\   
          & 24 April 2006    & 0.9--1.1           & 7200          &      \\
          & 19, 21 June 2006 & 1.0--2.0, 1.0--2.0 & 28430$^d$     &      \\
          & 10 April 2007    & 0.4--1.0           & 28800         &      \\\tableline
IOK-3     & 14, 15 May 2005  & 0.5--0.7, 0.9--1.0 & 3600          & MOS-2\\    
          & 1 June 2005      & 0.6--0.8           & 5400          &      \\
Obj-4$^e$ & 14, 15 May 2005  & 0.5--0.7, 0.9--1.0 & 1800          & MOS-5\\   
Obj-5$^e$ & 14, 15 May 2005  & 0.5--0.7, 0.9--1.0 & 3600          & MOS-3\\\tableline
\end{tabular}
\tablenotetext{a}{All the spectra taken during the observation.}
\tablenotetext{b}{Out of all the spectra taken, 11 30-min exposures were used to obtain the final combined spectrum.}
\tablenotetext{c}{Out of all the spectra taken, 22 30-min exposures were used to obtain the final combined spectrum.}
\tablenotetext{d}{We did not use data taken at this night because of their low quality (bad seeing).}
\tablenotetext{e}{Not identified yet.  The data analyses are in progress.}
%% Any table notes must follow the \end{tabular} command.
\end{center}
\end{table}

\clearpage

\begin{table}
\begin{center}
\footnotesize
\caption{Spectroscopic properties of the Ly$\alpha$ emission of the $z=6.96$ LAE\label{Spec-Property}}
\begin{tabular}{lccccccc}\tableline\tableline
Object & $z$  & $F(\rm Ly\alpha)$               & $L(\rm Ly\alpha)$    & $SFR(\rm Ly\alpha)$  & FWHM & Sw & S/N\\
       &      & ($10^{-17}$erg s$^{-1}$ cm$^{-2}$) & ($10^{43}$erg s$^{-1}$) & (M$_{\odot}$ yr$^{-1}$) & (\AA) (km s$^{-1}$) & (\AA) & \\\tableline
IOK-1  & 6.96 & 2.00                               & 1.13                    & 10.24                   & 13 ~~403 & $9.46\pm0.39$ & 5.5      \\\tableline
%%IOK-2  & 7.02$^a$ & 0.91                           & 0.52                    & 4.73                    & 14 ~~431 & $6.61\pm1.51$ & (2.3)$^b$\\\tableline
\end{tabular}
%%\tablenotetext{a}{Note that the spectroscopic identification of IOK-2 is unsecure and thus tentative.}
%%\tablenotetext{b}{The S/N ratio was measured within the gap of OH lines (See Figure \ref{MS_IOK2_Spec}).}
%% Any table notes must follow the \end{tabular} command.
\end{center}
\end{table}

\clearpage

\begin{table}
\begin{center}
\caption{Spectroscopic properties of the UV continuum of the $z=6.96$ LAE\label{Spec-Property_UV}}
\begin{tabular}{lccc}\tableline\tableline
Object & $z$      & $L_{\nu}(\rm UV)$               & $SFR(\rm UV)$\\
       &          & ($10^{29}$erg s$^{-1}$ Hz$^{-1}$)  & (M$_{\odot}$ yr$^{-1}$)\\\tableline
IOK-1  & 6.96     & 2.58                               & 36.1\\\tableline
%%IOK-2  & 7.02$^a$ & 4.53                               & 63.4\\\tableline
\end{tabular}
%%\tablenotetext{a}{Note that the spectroscopic identification of IOK-2 is unsecure and thus tentative.}
%% Any table notes must follow the \end{tabular} command.
\end{center}
\end{table}

\clearpage

\begin{table}
\begin{center}
\caption{The neutral IGM fractions obtained by several independent methods\label{Neutral_Frction}}
\begin{tabular}{lcccc}\tableline\tableline
                    &\multicolumn{4}{c}{Neutral fractions $x_{\rm HI}$} \\
Method              & $z\sim 6$  & $z\sim6.3$    & $z=6.6$    & $z=7.0 $   \\\tableline
(1) Quasar GP test$^a$  & 0.01--0.04 & ---           & ---        & ---        \\
(2) Gamma Ray Burst$^b$ & ---        & $<0.17$--0.60 & ---        & ---        \\
(3) Ly$\alpha$ LF$^c$  & ---        & ---           & $<0.45$    & ---        \\
(4) Ly$\alpha$ LF and LBG UVLF$^d$ & ---   & ---           & $\sim0.12$--0.42 & $\sim0.12$--0.54\\
(5) Model and observed Ly$\alpha$ LFs$^f$ & --- & ---    & $\sim0.24$--0.36 & $\sim0.32$--0.64 \\\tableline
\end{tabular}
\tablenotetext{}{(1) \citet{Fan06}.}
\tablenotetext{}{(2) \citet{Totani06}.}
\tablenotetext{}{(3) \citet{kashik06}.}
\tablenotetext{}{(4) This study and \citet{Yoshida06}, \citet{06BI} and \citet{Bouw06}.}
\tablenotetext{}{(5) This study and \citet[K07 model]{ktn07}.}
%% Any table notes must follow the \end{tabular} command.
\end{center}
\end{table}

\clearpage


%%figure 1

%%figure 2

%%figure 3

%%figure 4

%%figure 5

%%figure 6

%%figure 7

%%figure 8

%%figure 9

%%figure 10

%% The following command ends your manuscript. LaTeX will ignore any text

\end{document}